\documentclass[letterpaper]{article}
\usepackage{verbatim}
\usepackage{moreverb}
\usepackage{url}
\usepackage{amsmath, amsthm, amssymb}
\usepackage{color}
\usepackage{appendix}
\usepackage{graphicx}
\usepackage{subfig}
\usepackage[colorlinks=true,bookmarks=true,pdfauthor={Vadim Zaliva, lord@crocodile.org},
            pdftitle={Applying static code analysis to firewall policies for the purpose of anomaly detection},
            pdftex]{hyperref}
\usepackage{listings}
\usepackage{setspace} 
\usepackage{amsfonts}
\newtheorem{theo:mcsiunique}{Theorem}
\author{Vadim Zaliva, lord@crocodile.org}
\date{2009}
\title{Applying static code analysis to firewall policies for the purpose of anomaly detection}

\begin{document}

\maketitle

\pagenumbering{roman}
\pagestyle{empty}

\begin{abstract} 
Treating modern firewall policy languages as imperative, special
purpose programming languages, in this article we will try to apply
\textit{static code analysis} techniques for the purpose of anomaly
detection.

We will first abstract a policy in common firewall policy language
into an intermediate language, and then we will try to apply 
anomaly detection algorithms to it.

The contributions made by this work are:

\begin{enumerate}
\item An analysis of various control flow instructions in popular
  firewall policy languages
\item Introduction of an intermediate firewall policy language, with
  emphasis on control flow constructs.
\item Application of \textit{Static Code Analysis} to detect anomalies
  in firewall policy, expressed in intermediate firewall policy language.
\item Sample implementation of \textit{Static Code Analysis} of firewall
  policies, expressed in our abstract language using Datalog language.
\end{enumerate}
\end{abstract}

\newpage
\tableofcontents

\newpage
\listoffigures

\newpage\section{Introduction}
\pagenumbering{arabic}

Computer firewalls are widely used for security policy enforcement and
access control. They are used to block unauthorized network
access. This is usually done by partitioning a network into
\textit{security domains} and defining access rules on the boundaries
of these domains. The simplest case is when just two domains are
defined: LAN (Local Area Network) and WAN (Wide Area Network). In more
complex cases one might have domains for different departments within
an organization as well as different parts of the Internet (for
example different geographic regions). A firewall could be implemented
either as hardware -- a special device -- or as software running on
top of a general purpose operating system. Usually an organization
deploys more than one firewall and their policies must be coordinated
to provide a consistent level of security.

Firewall configuration, which mainly consists of writing
\textit{policy} (also sometimes reffered to as a \textit{rule set}) is an
exacting task, usually done by a human. Given the complexity of some
policies (hundreds, or even thoudands of rules) and human
predisposition to err, misconfigurations often occur. In a quantitive
study performed by Wool\cite{wool2004qsf}, 37 firewall rule
set were examined, collected from organizations in various market segments. He
discovered that all of them were misconfigured, most in multiple
places. The implications of his study are trully alarming: a firewall
misconfiguration could expose private customer information (including
medical records), sensitive business information, lead to financial
loss and in some cases even impact the life and safety of people
relying on the security of the computer system.

An example of misconfiguration is when a firewall could be managed
from an insecure location (any machine outside the network
perimeter). Usually access to firewall management interfaces is
limited to a secure domain inside the organization. The simplest form
of this misconfiguration is when a rule limiting such access is not
added to the policy. A novice firewall administrator can easily make
such a mistake. Another scenario in which a misconfiguration may occur
is when such a rule was added, but was overriden by another, usally
more generic rule. This is a mistake that even a more experienced
firewall adminstrator may make. At a cursory examination the policy
looks good, as the rule is present. However it will require a deeper
examination to discover that the rule does not have any effect.

The task of automatic discovery potential of firewall configuration errors
is called \textit{firewall anomaly detection}. There is some existing
research in this area summarized in
Section~\ref{subsec:anomalydetection}. In this article we attempt to
advance research in this area by applying \textit{static
  code analysis} to firewall policies.

\textit{Static code analysis} allows us to analyze and predict program
behavior without actually executing it. It is commonly used for
performance optimization and error detection.

In this section we will start by introducing basic firewall and
\textit{static code analysis} concepts.

\subsection{Packet Filtering}
Packet filtering is a core functionality of network firewalls. The
main idea is that the firewall resides on a network \textit{Node}
(\textit{Host} or \textit{Router}) and inspects all network
traffic. Inspection is performed in accordance with network security
policy (which we will discuss in detail later). Based on this policy,
the firewall makes a decision regarding what action to perform on a
given packet. The most commonly performed actions are:

\begin{description}
\item[ACCEPT] - the packet is permitted to pass through
\item[DENY] - the packet is silently dropped
\end{description}

Some firewalls allow additional actions, which do not necessarily
affect the packet's traversal of the firewall, but are invoked for
side effects. Common examples are logging and setting the values of some
\textit{variables} (which could be checked later in other policy rules).

Most modern firewalls also support actions, which affect the control flow
of packet processing through firewall rules.

Issues related to the low-level implementation of categorizing packets
and the algorithms for doing this efficiently are commonly referred to
as the \emph{packet classification problem}\cite{gupta2001apc}. This
problem mostly deals with performance and resource usage constraints.

Here is an example of a simple firewall policy. This policy permits
all incoming traffic on interface $dc0$ from local network
$192.168.0.0/24$ to host with IP address $192.168.0.1$. It also allows
the return traffic to pass.

\begin{verbatim}
pass in  on dc0 from 192.168.0.0/24 to 192.168.0.1
pass out on dc0 from 192.168.0.1 to 192.168.0.0/24
\end{verbatim}

\subsection{Firewall Policy}

The firewall's behavior is controlled by the \textit{Policy}. A policy
consists of \textit{Rules} (in the context of packet routing they are also
often referred to as \textit{Filters}). Each rule consist of a
\emph{condition} and an \emph{action}. Conditions describe the
criteria used to match individual packets. Actions describe the
activity to be performed if matches have been made. 

Basic conditions consist of tests, which match individual fields of the
packet such as source address, destination address, packet type,
etc. In the case of \emph{stateful inspection}, connection-related
variables like connection state (\emph{established}, \emph{related}, or
\emph{new}) can be checked. Finally, system state variables
such as current time of day, CPU load, or system-wide configuration
parameters can be taken into account.

The sequence of rules processing differs significantly between various
firewall implementations. There are two common matching strategies:

\emph{single trigger} processing means that an action of the first
matching rule will be performed.

\emph{multi-trigger} processing means that all rules will be matched and
an action from the last matching rule will be performed.

Here is a typical example of multi-trigger policy for \emph{pf}
firewall platform:

\begin{verbatim}
block in all
...
pass in on dc0 from any to 192.168.1.0/24 port 22
\end{verbatim}

In this example the first rule blocks all incoming traffic. This is
sometimes called a \textit{default deny} approach. The next rule
allows incoming traffic on port $22$ to LAN subnet
$192.168.1.0/24$. As an incoming packet destined for port 22 passes
the firewall, it will match the first rule and will be marked to be
dropped. However, since \emph{pf} by default is using multi-trigger
strategy, it will continue to try to match the packet to other
rules. It will match against the last rule and this time the action
will be changed to \emph{pass}. At the end of the policy, this last
action will take effect and the packet will be allowed to pass
through.

Some firewalls like \emph{ipfilter} support \emph{multi-trigger} strategy
by default, but allow individual rules to specify a \emph{quick} option,
which signifies that no further processing should be done on a matched
packet.

Let us try to express the previous example using single trigger
strategy. We will again use \emph{pf} syntax, but now we will use the
\emph{quick} keyword on all rules to enforce single trigger
processing. The resulting policy would be:

\begin{verbatim}
pass in quick on dc0 from any to 192.168.1.0/24 port 22
...
block in quick all
\end{verbatim}

Now, the incoming packet destined for port $22$ will match the first rule
and will be immediately allowed to pass. No other rules will be
considered. All incoming packets which have not matched the first rule
will match the last one and will be denied.

In addition to the \emph{single trigger} or \emph{multi-trigger}
control flow models, most popular modern firewall platforms support
more complex control flow models, with statements allowing conditional
or unconditional branching, early termination, sub-routine calls, etc.

\subsection{Static Code Analysis}

\textit{Static Code Analysis} is defined as:

\begin{quotation}
  ``Program analysis offers static compile-time techniques for
  predicting safe and computable approximations to the set of values
  or behaviors arising dynamically at run-time when executing a
  program on a computer.''\cite{nielson1999ppa}
\end{quotation}

We will concentrate on the type of static code analysis termed
\textit{data flow analysis}, first introduced in \cite{512945}.

In data flow analysis, the program is split into elementary blocks
which are organized into a directed graph. This is called a
\textit{control flow graph (CFG)}. Nodes are elementary blocks, and
edges indicate how control can pass between them. For each block we
can define a set of equations, describing information at the exit of a
node, as related to information upon entry to the node. 

Let us consider a simple example of \textit{Reaching Definitions
  Analysis}, which is defined in \cite{nielson1999ppa} as follows:

\begin{quotation}
  ``For each program point, which assignments \textit{may} have been
  made and not overwritten, when program execution reaches this point
  along some path.''
\end{quotation}

Now for each label $l$ we can write two sets of reaching defintions
(abbreviated as $RD$): $RD_{entry}(l)$ and $RD_{exit}(l)$, each of
them defining a list of tuples in the form of $(variable, label)$
defining what variables were assigned at what labels. The special
label $?$ could be used to indicate the possibility of an
uninitialized variable.

For example if at label 2 variable $z$ is assigned a new value we can
write: $RD_{exit}(2) = (RD_{entry}(2)\setminus\{(z,l) | l \in L\})
\cup {(z,2)}$ where $L$ is set of all program labels.

In general, $RD_{entry}(l)=RD_{exit}(l_1) \cup \ldots \cup
RD_{exit}(l_n)$ if $l_1, \ldots, l_n$ are labels from which contol may
reach $l$. For \textit{initial label} all variables are associated
with $?$ label.

Using these equations we can reason about properties of the program at
the block boundaries. The standard way to do this is to solve the
system until it reaches \textit{fixpoint}.

We must distinguish between two types of data flow analyses. In
\textit{forward data flow analysis} values are propagated in a control
flow graph in the direction of the control flow. In \textit{backward
  data flow analysis}, the values are propagated in the opposite
direction of control flow.

Another useful type of static code analysis is called \textit{control
  flow analysis}, which deals with the calculation of control flow graphs
(which blocks lead to which).

\subsection*{Article Organization}
The rest of this article is organized as follows:

First, in Section~\ref{sec:formalmodel} we will briefly discuss the
current state of research in Firewall Policy modeling and analysis.

Then, in Section~\ref{sec:concrete} we will review control flow models
and control flow statements in several existing firewall platforms,
concentrating on the few most popular ones.

In Section~\ref{sec:abstract-lang} we define an \textit{intermediate
  rule language} for policy rule processing which includes support for
all control flow constructs mentioned in previous sections. We will
show how all firewall languages described in Section~\ref{sec:concrete}
can be converted into this new intermediate language.

In Section~\ref{sec:analysis} we will show how \textit{static code
  analysis} can be applied to a firewall policy as expressed in
\textit{intermediate language}.

And finally, in Section~\ref{sec:implementaion} we will present our
implementation of policy analyzer using techniques described in
Section~\ref{sec:analysis}. It will be followed in
Section~\ref{sec:examples} by some usage examples of our
analyzer.

\newpage\section{Firewall Policy Modeling}
\label{sec:formalmodel}

In this section we discuss approaches used in modeling firewall
policies and briefly review the current state of research in this
area.

Many researchers assign to the policy a declarative semantics,
treating it as a set of tuples (e.g. \cite{alshaer2004dpa},
\cite{cuppens2005dar}, \cite{capretta2007fcc}, \cite{liu2005crd}).
Each tuple contains conditions used to match various packet fields and
actions. For example, Ehab S. Al-Shaer and Hazem H. Hamed
\cite{alshaer2003fpa},\cite{alshaer2004dpa} use a fixed rule structure,
called a ``5-tuple filter'':
\textit{(order,protocol,src\_ip,src\_port,dst\_ip,dst\_port,action)}.

\begin{figure}[htp]
\centering
\includegraphics[width=4in]{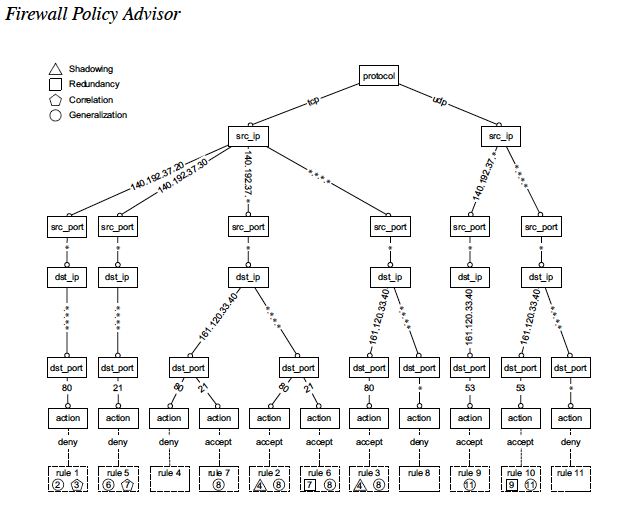}
\caption{Example of a policy representation as a tree}
\label{fig:policyadvisortree}
\end{figure}

In order to formally model firewall policy, these researchers start by
defining pairwise relationship between rules in the policy:
``completely disjoined'', ``exactly matched'', ``inclusively
matched'', ``partially disjoined'', and ``correlated''. Next Al-Shaer
and Hamed prove that these relationships are distinct and that their
union represents the universal set of relations between any two
k-tuple filters in a firewall policy. The policy is represented as a
single-rooted tree, where each node represents a field of a filtering
rule and each branch at this node represents a possible value of the
associated tree. An example of such a tree taken from
\cite{alshaer2003fpa} is shown in Figure~\ref{fig:policyadvisortree}.
Thus each path in the tree (starting from root) represents a policy
rule. Each branch has at the end an action leaf, which shows the
action which should be taken. The dotted box at the bottom lists rules
associated with a given branch. Normally each branch should have only
one rule associated with because ideally each packet should be
processed by a sigle rule. If more than one rule is associated it
represents an anomaly. For example rule 8, in addition to having its
own branch, also appears on branches for rules 2,3,6, and 7. This
happens because the rule is a superset of any of these other rules,
and packet matching any of them will match rule 8 as well.

Hari et al. \cite{hari2000dar} considers a much simpler packet
filtering model, where each filter is k-tuple $(F[1],F[2], \ldots,
F[k])$ and where each field $F[i]$ is a prefix bit string. This model
could be used not only in firewalls, but also for routing. Note that
all matching is done only by matching prefix bit strings. However, as
shown in \cite{srinivasan1998fas}, it is always possible to represent
a sub-range of $[0,2^k]$ as at most $2k$ prefixes. This allows us to
convert from range-based policy rule representation to
prefix-based. Prefix-based policy rule representation has a useful
property, on which Hari et al. base their algorithms:

\begin{quotation}
  ``If filter fields are prefix fields, then each field of a filter is
  either a strict subset of, or equal to, or a strict super-set of, or
  completely disjoint from the corresponding field in any other
  filter. In other words, it is not possible to have partial overlaps
  of fields. Partial overlaps can only occur when the fields are
  arbitrary ranges, not prefixes.''\cite{hari2000dar}
\end{quotation}

Using this property, the authors propose to solve a filter conflict problem
by formulating it as a cycle elimination problem in a directed graph.

There are some efforts related to the analysis of firewall policies
using machine reasoning techniques. In particular, 
\cite{eronen2001esa} describes an Expert System built using Constraint
Logic Programming (CLP). Considering each rule as 6-tuple or as ranges
along with action taken (``permit'' or ``deny''), the system
represents them as constraints on the 6-dimensional packet space. Each
rule is a 6-dimensional hypercube.

Capretta et al. in \cite{capretta2007fcc} use the
Coq\cite{bertot2004itp} proof assistant to detect conflicts in
firewall policies. Their ``conflict'' definition is two rules for
which there exists a request to which they give an opposite action (only
\emph{ACCEPT} and \emph{DENY} actions are considered). Then they
formally prove soundness and completeness to establish the
correctness of their algorithm.

Another approach to policy modeling is using geometric interpretation.
For example, Eppstein\cite{eppstein2001ipf} suggests that each rule
could be represented as a collection of $d$-dimensional ranges
$[l_i^1,r_i^1] \times \ldots \times [l_i^d,r_i^d]$, an action $A_i$
and priority $p_i$. Similarly, each packet can be viewed as a
$d$-dimensional vector of values $P=[P_1, \ldots, P_d]$. In IP network
terms, each dimension could correspond to an IP packet field. Thus the
range $[l_i^n,r_i^n]$ allows us to check if the value of an IP packet field
number $n$ falls in range $[l_i,r_i]$. A filter $i$ applies to packet
$P$ if $P_j \in [l_i^j,r_i^j]$. Epspstein proceeds to formally define
\emph{packet classification} and \emph{filter conflict detection}
problems using this geometrical abstraction and suggests algorithms
for solving them.

The \emph{multidimensional range searching} problem from computational
geometry is related to filter conflict detection. Multiple
algorithms exist to solve this problem, surveyed in \cite{lin1998cdb}.
In particular, as mentioned in \cite{hari2000dar}, Edelsbrunner
\cite{edelsbrunner1983nar} has proposed an algorithm which, in the
worst case, can solve this problem in $O((log(N))^{2k-1}+R)$ where $N$
is the number of \emph{k}-dimensional rectangle boxes and $R$ is
number of boxes intersecting the query box. However as Hari et. al
conclude in \cite{hari2000dar}, there are several known problems with the
geometric interpretation approach:

The first problem is that this intepretation treats one filter which is
fully contained within another filter as the intersection of their
rectangles. However this does not indicate a conflict, since the
contained filter is more specific. Thus not every detected
intersection indicates a conflict (some intersections are
false-positives).

The second problem relates to time and space bounds. Quick
estimation from \cite{hari2000dar} concludes:

\begin{quotation}
  ``... even for modest values of $N$ and $k$ , the worst-case time
  and space bound guaranteed by this data structure are hopelessly
  bad. For instance, when $N=10,000$ and $k=4$, the algorithm
  guarantees a worst-case search cost of $13^7=62748517$ , meaning
  that it is no better than a linear search through the
  filters.''\cite{hari2000dar}
\end{quotation}

Guttman\cite{guttman2005ran} et al. describe a group of network security
related problems and modeling frameworks that lead to their solutions:
\begin{quote} We focus the modeling work on representing behavior as a
  function of configurations, and predicting the consequences of
  interactions among differently configured
  devices.\cite{guttman2005ran}
\end{quote}

While Guttman et al. cover both packet filtering firewalls and IPSec
gateways, Uribe et al.\cite{uribe2007aaf} build upon their work,
extending it by including specifications and requirements for Network
Intrusion Detection Systems (NIDSs).

All these approaches are based on declarative interpretation of
firewall policy. Another approach is to assume an imperative
semantics, treating firewall policy as a set of statements, the execution
of which is controlled by a control flow.

In practice, most firewall implementations support imperative rather
than declarative semantics, so this second approach has more practical
applications.  However, most researchers deal with the simplest form
of control flow when analyzing firewall policies, where the policy is
a list of sequentially applied rules. Most modern popular firewall
platforms support more complex control flow models, with statements
allowing conditional or unconditional branching, early termination,
sub-routine calls, etc.

Yuan et al. in FIREMAN\cite{yuan2006ftf} are some of the few
researchers who go beyond a simple linear policy model and consider
what they call a \emph{Complex Chain Model}, covering a more complex
policy organization similar to that implemented in the popular Linux
firewall Netfilter. They also introduce the notion of an \emph{ACL
  Graph}, formed by a combination of multiple ACLs (access control
lists) across the trajectory of the packet. Using this graph they
provide some analysis of anomalies in distributed firewall
configuration. Their approach is similar to ours (they also use a form
of static code analysis) but they use a slighly different model of
firewall policy. Their model, apparently inspired by \emph{iptables},
is organized around \textit{chains} and the only branching instruction
supported is a ``calling chain''. Our approach is more generic and
allows us to accomodate branching models found in other firewall
products, like \emph{pf} or \emph{ipfw}. To reason about rules, they
use set operations on ACL. Operations like $\subseteq$ and $\cap$ have
to be defined for all packet field types could be computationally
heavy, compared to our symbolic approach using \textit{MCSI}.

\subsection{Anomaly Detection}
\label{subsec:anomalydetection}

Eppstein et al.\cite{eppstein2001ipf} define the \emph{filter conflict
  detection problem} as a way to detect when two or more filters
(rules) applied to a packet specify conflicting actions. For example
some let a packet pass through while others reject it. The presence of
such rules in a policy could indicate an error and could lead to
firewall misconfiguration.

Some studies\cite{gupta2000arl} show that 15\% of rules in real-life
policies might be redundant. A more formal definition of shadowing and
redundancy, quoted from \cite{cuppens2005dar} is as follows:

\begin{quotation}
  ``\textbf{Definition 1.1} \textit{Let R be a set of filtering
    rules. Then R has shadowing iff there exists at least one
    filtering rule, $R_i$ in R, which never applies because all the
    packets that $R_i$ may match, are previously matched by another rule,
    or combination of rules, with higher priority in order.}
    
  \textbf{Definition 1.2} \textit{Let R be a set of filtering
    rules. Then R has redundancy iff there exists at least one
    filtering rule, $R_i$ in R, such that the following conditions hold:
    (1) $R_i$ is not shadowed by any other rule; (2) when removing $R_i$
    from R, the filtering result does not change.}''
\end{quotation}

In \cite{cuppens2005dar}, the authors present (with formal proofs of
correctness) algorithms for shadowed and redundant rule detection and
removal. However, their algorithms use a simplified firewall model
with only a \emph{single trigger} processing strategy and just two
possible actions: \emph{ACCEPT} and \emph{DENY}. The authors do not go
into exact semantics of rule condition matching, treating them as a
conjunctive set of opaque condition attributes such that $condition_i
= A_1 \wedge A_2 \wedge \ldots \wedge A_p$ – $p$ being the number of
condition attributes of the given set of filtering rules.

In \cite{alshaer2003fpa} the authors identify four firewall policy
anomalies:

\paragraph{Shadowing anomaly} A rule is shadowed when a previous rule
matches all the packets that match this rule, such that the shadowed
rule will never be activated.

\paragraph{Correlation anomaly} Two rules are correlated if the first
rule in order matches some packets that match the second rule and the
second rule matches some packets that match the first rule

\paragraph{Generalization anomaly} A rule is a generalization of
another rule if this general rule can match all the packets that match
a specific rule that precedes it.

\paragraph{Redundancy anomaly} A redundant rule performs the same
action on the same packets as another rule such that if the redundant
rule is removed, the security policy will not be affected.

Al-Shaer and Hamed proceed to provide some algorithms to detect
any of these anomalies.

In \cite{alshaer2004dpa}, they extend their anomaly-detection
algorithms to include configuration, consisting of multiple firewalls. They
provide format definitions of various \emph{Inter-Firewall Anomalies}
and propose algorithms for their detection.

Baboescu \cite{baboescu2003fas} suggests an optimized conflict detection
algorithm which, while based on a known Bit Vector approach, 
shows an order of magnitude of improvement compared to previous work.

Qian et al. in \cite{qian2001afa} introduce a framework which includes
algorithms, allowing it to:

\begin{quote}
  ``detect and remove redundant rules, discover and repair inconsistent
  rules, merge overlapping or adjacent rules, map an ACL with complex
  interleaving permit/deny rules to a more readable form consisting of
  all permits or denies, and finally compute a meta-ACL profile based
  on all ACLs along a network path.''\cite{qian2001afa}
\end{quote}

They present a set of formal \emph{rule relation} definitions:
``intersect'', ``contain'', ``overlap'', ``disjoint'', ``adjacent'',
``inconsistent'' and ``redundant''.

Gouda and Liu in \cite{liu2005crd} analyze the rule redundancy problem.
They introduce the notion of \emph{upward redundant rules} and
\emph{downward redundant rules} (with formal definition). They offer
algorithms for the identification of redundant rules using a \emph{firewall
  decision tree}.

\newpage\section{Control flow models in existing firewall platforms}
\label{sec:concrete}

In this section we will review control flow models and control flow
statements in existing firewall platforms. We will concentrate on the
few most popular ones.

\subsection{Netfilter}

Netfilter\cite{netfilter} is a very popular firewall platform, and is
the default firewall platform for most Linux distributions.  In
Netfilter, policy rules are organized into \textit{chains}. There are
three built-in chains: \emph{INPUT}, \emph{OUTPUT}, and
\emph{FORWARD}.  In addition, there exists the possibility for
additional user-defined chains.

Built-in chains have a \textit{policy} which determines the default
action for this chain -- what happens if the packet reached the
end of the chain and no action has been taken on it so far. The
policy could be either \emph{ACCEPT} or \emph{DROP}. User-defined
chains all have an implicit \emph{RETURN} policy which cannot be
changed by the user.

Each rule can specify as an action a \emph{JUMP} to another
chain. This means that if rule conditions match, the packet will
continue its traversal starting from the beginning of the specified
chain. If the packet has reached the end of the called chain, or the
\emph{RETURN} action was triggered explicitly (by a rule) it will
continue processing from the next rule of the caller chain. Upon return from
built-in chains, a chain policy action is executed. For user-defined
chains control returns to the caller.

Instead of \emph{JUMP}, the user can specify a \emph{GOTO} action in order to
branch to another chain. When \emph{RETURN} is encountered in the other
chain, the control will return not to this chain, but to the chain which
has called this one via \emph{JUMP}.

Both \emph{JUMP} and \emph{GOTO} can call any user-defined chain,
except the one which contains this rule.

Loops are not permitted with \emph{JUMP} and \emph{GOTO}. Such loops
are detected by an \emph{iptables} command and are reported with the
somewhat cryptic message ``iptables: Too many levels of symbolic
links''. Loops detection in the current version (1.4.0) seems to be
very rough: it basically forbids any JUMP/GOTO loops regardless of the
conditions under which they occur. For example, the following policy
is not accepted, because it is considered to contain a loop:

\begin{verbatim}
iptables -A INPUT -p icmp -d 38.99.76.17 -g test0 
iptables -A test0 -p icmp -d 38.99.76.18 -j test1 
iptables -A test1 -p icmp -d 38.99.76.19 -j test2 
iptables -A test2 -p icmp -d 38.99.76.20 -g test0 
\end{verbatim}

The \emph{SET-MARK} action\footnote{Implemented in \emph{mark} module}
module can be used to associate values (\textit{marks}) with a
packet while it is being processed. These marks can be checked later
using a \emph{MARK} clause in filtering
specification\footnote{Marks can be set only in the \emph{mangle table} but
  can be checked in any other tables. Although it is not recommended
  to use mangling table for filtering, it is possible to do so. For
  completeness we assume that marks may be set and checked in any
  chain}. Only one mark value can be associated with the
packet. Setting a new mark value will replace the old one.

Mark checks may be performed with \emph{MARK} filtering
specification by comparing its value to a given constant, after
applying an optional \textit{mask} using a bitwise \emph{AND} operation.

The following is a simplified grammar of \textit{iptables} policy language:

\begin{verbatim}
chain ::= (rule)*
rule ::= filtering-spec target
filtering-spec := mark-filtering-spec | ...
mark-filtering-spec := MARK mark-number ["/" mark-mask]
target ::= ACCEPT | DROP | RETURN | 
           JUMP chain | GOTO chain | 
           SET-MARK mark-number
mark-number := 0..(2E32-1)
mark-mask := 0..(2E32-1)
\end{verbatim}

Here is an example of a simple \emph{netfilter} policy:

\begin{verbatim}
iptables --policy INPUT DROP
iptables --policy OUTPUT DROP
iptables --policy FORWARD DROP
iptables -A INPUT -m state --state ESTABLISHED,RELATED -j ACCEPT
iptables -A OUTPUT -m state --state NEW,ESTABLISHED,RELATED -j ACCEPT
iptables -A INPUT -p tcp -s 192.168.1.100 --dport 22 -m state \
            --state NEW -j ACCEPT
iptables -A INPUT -j DROP
\end{verbatim}

The processing model is fairly simple. The processing starts from one
of the built-in chains, and proceeds sequentially until either an \emph{ACCEPT}
or \emph{DROP} action is triggered, or control is passed to another
chain. This model is very similar to the execution model of instructions
in a CPU. Rules can call other chains (as subroutines) or branch to them
(similar to the infamous \emph{GOTO} instruction in programming
languages).

\subsection{PF}

PF\cite{pf} (stands for ``packet filter'') is another popular firewall
platform, and the default firewall for the OpenBSD system.  The main
difference between PF and Netfilter is that by default, unless the
\emph{quick} option is specified, it is using \textit{multi-trigger}
matching strategy. Thus the last matched target will be used to
determine what to do with the packet. The \emph{quick} option causes
it use \textit{single-trigger} strategy for current rule.

In PF there is a notion of \textit{anchors}. An \textit{anchor} is a
set of rules which could be invoked at any point of the policy using
the \emph{anchor} instruction. Anchors could be defined independently or
as a part of the main ruleset (inline anchors). Anchors can be
nested.

The following is a simplified grammar of the \textit{pf} policy language:

\begin{verbatim}
ruleset := (rule | anchor)*
rule : = filtering-rule | anchor-rule
filtering-rule := filtering-spec target [quick]
anchor-rule := ANCHOR anchor-name(/anchor-name)*
target ::= ACCEPT | DROP | TAG tagname
anchor ::= [anchor-name] ruleset
\end{verbatim}

Here is an example of a simple \textit{pf} firewall policy:

\begin{verbatim}
anchor "goodguys" { 
   pass in proto tcp from 192.168.2.3 to port 22
}
...
anchor goodguys
pass in  on dc0 from 192.168.0.0/24 to 192.168.0.1
pass out on dc0 from 192.168.0.1 to 192.168.0.0/24
\end{verbatim}

The default target for the main ruleset is \emph{PASS}, which will be used if
the packet has not matched any rules.

Upon a match, rules can \textit{tag} a packet. Only one tag can be
assigned to the packet at a time, and once it is assigned, it 
cannot be removed. Subsequent rules can check for the tag presence.

When anchors are defined inline (either via curly brackets syntax, or via
the \emph{LOAD ANCHOR} command) they are also invoked at the place of
definition.

Informally the processing model could be described as follows: the
rules are processed one by one. The last matching action is remembered
and will be used when the end of the policy is reached
(\textit{multi-trigger} strategy). A rule with a quick option will
cause its action to take effect immediately (\textit{single-trigger}
strategy). Anchors are just named blocks of rules which can be invoked
at some points of the policy, either unconditionally or based on the
results of the evaluation of a packet matching criteria. Packets can
be tagged with a single named tag which is sticky. Rules can check for
the presence of a particular tag.

\subsection{IPFW}

IPFW (also known as IPFIREWALL)\cite{FBSDHB:IPFW,ipfw8} is a firewall
platform sponsored, authored, and maintained by the FreeBSD project. It
is also used as a firewall under MacOS. IPFW matches rules
sequentially, stopping at the first matching rule. Rules are numbered
from 1 to 65535. If no rules are matched, the packet is discarded by the
default rule with number 65535. Each rule belongs to exactly one set
from 0 to 31 with 0 being default. Some sets could be enabled or
disabled at any given time, except set 31 which is always
enabled. Sets are just a way for the firewall administrator to organize
policy rules. They are not part of the rule language.

A packet can have zero or more \textit{tags} associated with it. Tags
are identified by numbers in the range $[1..65534]$. Tags can be set
or unset conditionally (using tag or untag commands). Rule matching
can check for tag presence (using the \textit{tagged} rule option).

One action affecting the rule application sequence is
\emph{SKIPTO}. This action makes a firewall skip all rules with
numbers less than a specified amount.

The following is a simplified grammar of the \textit{ipfw} policy language:

\begin{verbatim}
ruleset := (rule)*
rule : = rule-number action filtering-spec
rule-number := 0..65535
set-number := 0..31
tag-number := 0..65535
action ::= ALLOW | DENY | TAG tag-number | UNTAG tag-number | 
           SKIPTO rule-number 
\end{verbatim}

Here is an example of a simple \emph{ipfw} policy:

\begin{verbatim}
501 deny all from any to any frag
502 deny tcp from any to any established
503 allow tcp from any to any 80 out via tun0 setup keep-state
504 allow tcp from any to 192.0.2.11 53 out via tun0 setup keep-state
\end{verbatim}

The \emph{set-number} is specified outside of the policy file, as an
argument to \emph{ipfw} command when it is loaded.

One interesting type of rules are \textit{probabilistic rules}, a
rules which match packets with a given probability. For the purpose of
anomaly detection, we will treat them as normal rules (triggered with
a probability of 1).

Informally, a processing model can be described as follows: rules
are always processed in increasing order of their numbers. Rules
can set and check one of 65536 Boolean variables. Additionally, there
is a \emph{SKIPTO} instruction, (similar to \emph{GOTO} in programming
languages), except it is only allowed to jump in a forward direction.

\subsection{IPFilter}

IPFilter (also known as IPF)\cite{ipfilterhome,conoboy1911ifb} is
used on many firewall platforms, most notably FreeBSD, NetBSD and
SUN Solaris.

IPFilter, like PF, uses the last matched rule to make decisions on how
to handle the packet (also known as \textit{multi-trigger}
processing). Also, like PF it has a \textit{quick} option to make
decisions immediately.

The \emph{SKIP} action skips a given number of rules.

Rules can also be placed in \textit{groups}. The default group is
0. A \emph{HEAD} parameter in the rule indicates that if matched,
further execution should proceed with rules within this group, using
this rule action as default (if none matched). If the \textit{quick}
keyword was specified, after processing a group specified in the
\emph{HEAD} parameter, packet processing stops. If it was not
specified, the firewall will continue rule processing in the group which
was active when \emph{HEAD} was executed. \emph{HEAD} instructions
can be used within non-default groups as well, to represent more
levels of branching.

The following is a simplified grammar of \textit{IPFilter} policy language:

\begin{verbatim}
ruleset := (rule)*
rule := target filtering-spec [group-number] [quick]
group-number := 0..65535
target := skip-target | regular-target
skip-target := SKIP number-of-rules
regular-target := [HEAD tag-number] (PASS | BLOCK)
\end{verbatim}

Here is an example of a simple \emph{ipfilter} policy:

\begin{verbatim}
block in all
block out all
pass in from firewall to any
block in on le0 proto icmp all
pass in on le0 proto 4 all
block in on le0 from localhost to any
block out quick on xl1 all head 10
pass out quick proto tcp from any to 20.20.20.64/26 port = 80 group 10
block out on xl2 all
\end{verbatim}

Group invocation is similar to a subroutine call in programming
languages, the main difference being that the invocation rule always
specifies a default action. Additionally, in subroutines the
\emph{SKIP} instruction provides a conditional jump forward within a
group.

\newpage\section{Intermediate Rule Language}
\label{sec:abstract-lang}

We will now try to define a generalized \textit{abstract syntax} for
policy rule processing which includes support for all control flow
constructs mentioned in the previous sections of this paper. We will
show that all firewall languages described above can be converted into
this new intermediate language. Having such a unified rule language,
we can do policy analysis without being dependent on specific
platforms.

This will be a language to express firewall filtering policies.  A
\textit{policy} consists of \textit{rules}. Each rule consists of two
parts: \textit{filtering specification} and \textit{target
  specification}. The language will have an imperative semantics. The
policy, expressed as a program in this language, will be applied to
each packet and as a result will produce an outcome -- how this packet
should be handled.

\subsection{Filtering specification} 

A \textit{filtering specification} is a predicate, which is evaluated
during rule processing. If it evaluates to \emph{True}, the
\textit{target specification} is invoked.

For the purpose of this article, we will consider a \textit{filtering
  specification} to be a \emph{conjunction} of two predicates:
a \textit{static check} predicate and a \textit{dynamic check} predicate.

\paragraph{A \textit{static check}} predicate deals only with the
fields which do not change while an individual packet is matched
towards a policy. For example, a static check could examine packet
fields, firewall settings, etc. It should be noted that
\textit{stateful packet inspection}\cite{wiki:spi} is done here,
because a state cannot change during single packet processing.

In this article we will consider simplified static check specification
in the 5-tuple form: \emph{(src\_addr, src\_port, dst\_addr,
  dst\_port, protocol)}. All fields are \textit{intervals}, and the
corresponding fields of the packet are checked to see if they belong
to the following intervals:

\begin{description}
\item[src\_addr] Source address interval
\item[src\_port] Source port interval (for TCP and UDP)
\item[dst\_addr] Destination address interval
\item[dst\_port] Destination port interval (for TCP and UDP)
\item[protocol] IP Protocol number interval
\end{description}

Address fields (\emph{src\_addr} and \emph{dst\_addr}) are intervals
of IP addresses. For now, we will consider only IPv4 addresses, so the
values will be in the range of 0 to $2^{32}-1$ (inclusive).

Port ranges for \emph{src\_port} and \emph{dst\_port} are just numeric
intervals with values in the range of 0 to $2^{16}-1$ (inclusive).

Although a \textit{protocol} will usually be compared to a single
specific numeric value, we will represent this as checks towards an
interval (0 to $2^{8}-1$ inclusive), for consistency with other
fields.

Real firewalls can match some additional fields, like Data Link Layer
address or ICMP protocol type and code. Although we are not
considering them in this article, our methodology could easily be
extended to include more fields in the static check predicate.

\paragraph{A \textit{dynamic check}} predicate deals with values
which could be changed during packet inspection, for example by
\textit{target specifications} of previously matched rules. In our
language such values are stored in \textit{variables}. So, dynamic
checks are limited to testing values of such variables. See
Section~\ref{vars} for more details on variables.

\subsection{Target specification} 

A \textit{target specification} has an imperative semantics. It can
make a decision on how a packet should be processed. It can also
affect a sequence of rule executions as well as produce some
side-effects, like setting variables (which could be later examined by
the \textit{dynamic check} part of filtering specifications of rules,
executed after this one.)

Let us define target specifications. We will consider:

\begin{description}
\item[\emph{DROP}] denotes that the packet should not be allowed to
  pass through the firewall. It should be immediately silently dropped.
  No further rules should be evaluated.

\item[\emph{ACCEPT}] denotes that the packet should be immediately
  allowed to pass through the firewall. No further rules should be
  evaluated.

\item[\emph{CALL}] target means that the rule processing should
  proceed from the specified label until either packet is dropped,
  accepted, or the \emph{RETURN} action is invoked. \emph{CALL}
  instructions can be nested.

\item[\emph{RETURN}] target causes rule execution to proceed from
  instruction, immediately following the last \emph{CALL} instruction.
  If there was no \emph{CALL} instruction invoked, then the behavior
  of \emph{RETURN} is undefined.

\item[\emph{JUMP}] target means that rule processing should continue
  from a specified label.

\end{description}

During rule set processing either the \emph{ACCEPT} or \emph{DROP} action
should be triggered. If rule set processing is finished (the last rule
has been processed), and neither the \emph{ACCEPT} nor the \emph{DROP} action
has encountered the outcome of a packet, processing is undefined.

\subsection{Labels}
The rules are numbered. The label is a rule number. No two rules can
have same label. Some labels might not have rules associated with them.

\subsection{Variables} 
\label{vars}
In our language we will have a set of variables with a name and a
value. The name is a positive integer. The value is opaque (we do not
make any assumtion about value structure or type) and the only
operation allowed on it is an equality check, comparing it to a
literal or constant arithmetic expression (currently only the bitwise
\textit{and} operation between constants is permitted).

Unset variables are assumed to have a special value of
\emph{NIL}. Once the variable is set, it could be unset again by
assigning \emph{NIL} as a new value. A variable could be checked for
\emph{NIL} value comparing it to special \emph{NIL} literal.

\subsection{Abstract Syntax}
\label{subsec:abstract-lang-abstract-syntax}

An \textit{Abstract Syntax} ``specifies the set of trees that are
considered abstract representation of well formed documents in the
language''\cite{808259}.

We are using lower case letters for \textit{operators} and capital
letters for \textit{phyla} names. The abstract syntax for our
intermediate language is as follows:

\begin{lstlisting}[title={Atomic Operators}]
true
false
int
octet
long
varname
opaqvalue
nil
accept_target
drop_target
return_target
\end{lstlisting}

\begin{lstlisting}[title={Fixed Arity Operators},mathescape=true]
policy_def $\rightarrow$ POLICY
interval_def $\rightarrow$ PACKETFIELD FLAG PACKETFIELD FLAG
rule_def $\rightarrow$ LABEL STATIC_CHECK DYNAMIC_CHECK TARGET
static_check_def $\rightarrow$ FLAG INTERVALSET 
                               FLAG INTERVALSET
                               FLAG INTERVALSET
                               FLAG INTERVALSET
                               FLAG INTERVALSET
var_value_check_def $\rightarrow$ FLAG VAR VALUE
const_expr $\rightarrow$ CONST_VALUE
const_masked_expr $\rightarrow$ CONST_VALUE CONST_MASK
call_target $\rightarrow$ LABEL
jump_target $\rightarrow$ LABEL
var_set_target $\rightarrow$ VAR OPAQUE_CONST
\end{lstlisting}

\begin{lstlisting}[title={List Operators},mathescape=true]
ruleset_def $\rightarrow$ RULE ...
intervalset_def $\rightarrow$ INTERVAL ...
\end{lstlisting}

\begin{lstlisting}[title={Phyla}]
POLICY :: ruleset_def
TARGET :: accept_target drop_target 
          call_target jump_target return_target 
          var_set_target
VAR :: varname
VALUE :: nil opaqvalue const_expr const_masked_expr
OPAQUE_CONST :: opaqvalue
STATIC_CHECK :: static_check_def
DYNAMIC_CHECK :: var_value_check_def
FLAG :: true false
PACKETFIELD :: int
INTERVAL :: interval_def
INTERVALSET :: intervalset_def
CONST_VALUE :: int
CONST_MASK :: int
LABEL: long
\end{lstlisting}

\subsection{Sample Concrete Syntax}
\label{subsec:abstract-lang-concrete-syntax}
In this section we will define a simple concrete syntax which we will
use for examples in the rest of the document. The very same syntax
will be used in our proof of concept implementation.

The following is the syntax, expressed in EBNF notation\cite{scowen1993ebg}:

\begin{lstlisting}[caption={Sample Concrete Syntax},mathescape=true,
  breaklines=true,basicstyle=\small, language={}]
ruleset = {rule} ;
digit = "0" | "1" | "2" | "3" | "4" | "5" | "6" | "7" | "8" | "9" ;
ws = (" " | "\t" ), {" " | "\t" } ;
lf = ("\r" | "\n"), {"\r" | "\n"} ;
comment = "#", comment-text, lf ;
comment-text = (?Printable ASCII Characters? - ("\r" | "\n")) ;
opaque-value = "\'", (?Printable ASCII Characters? - ("\r" | "\n" | "\'")),
               "\'",;
number = digit, {digit} ;
var-name = "$\$$", number ;
rule = number  "if", filtering-spec, "then", target,  ";" ;
filtering-spec = (static-checks, "and", dynamic-check)
                 | static-checks
                 | dynamic-check
                 | "true" ;
dynamic-check = ["!"], var-name, "=", (opaque-value | const-expr | "nil") ;
addr_octet = digit, [digit [digit]] ;
const-expr = number, [ "&",  number] ;
target = action-target | branching-target | side-effects-target ;
action-target = "accept" | "drop" ;
branching-target = "call", number | "jump", number | "return" ;
side-effects-target = var-name, '=', "'", opaque-value, "'" ;
intrv_open = "[" | "(" ;
intrv_close = "]" | ")" ;
ipv4addr = addr_octet, ".", addr_octet, ".", addr_octet, ".", addr_octet ;
ipv4mask = addr_octet, ".", addr_octet, ".", addr_octet, ".", addr_octet ;
addr_interval = intrv_open,  ipv4addr,  ",",  ipv4addr,  intrv_close
                | ipv4addr, '/', number
                | ipv4addr, ':', ipv4mask;
port_interval = intrv_open,  number,  ",",  number,  intrv_close ;
proto_interval = intrv_open,  number,  ",",  number,  intrv_close ;
addr_interval_set =  "{", addr_interval, {",",  addr_interval},  "}"
port_interval_set = "{", port_interval, {",",  port_interval},  "}"
proto_interval_set = "{", proto_interval, {",",  proto_interval},  "}"
static-checks = ["!"], ("saddr", "in", (addr_interval | addr_interval_set))?,
                ["!"], ("sport", "in", (port_interval | port_interval_set))?,
                ["!"], ("daddr", "in", (addr_interval | addr_interval_set))?,
                ["!"], ("dport", "in", (port_interval | port_interval_set))?,
                ["!"], ("proto", "in", (proto_interval | proto_interval_set))? ;
\end{lstlisting}

\subsection{Mapping policy languages to Intermediate Policy language}

In this section we will show how policy languages of the concrete
firewalls we reviewed in Section~\ref{sec:concrete} could be
represented in our intermediate policy language. 

\subsubsection{Netfilter}

All rules in all chains are assigned non-overlapping ranges of
\textit{labels}. All rules within each chain have sequential labels.

\emph{ACCEPT}, \emph{DROP} and \emph{RETURN} are mapped to
corresponding intermediate language instructions.

\emph{JUMP} is mapped to \emph{CALL} with the chain name being mapped to
the label of it's first rule.

\emph{GOTO} is mapped to \emph{JUMP} with the chain name being mapped to
label of it's first rule.

The \textit{default policy} for built-in chains is mapped into an
additional rule, added at the end of each built-in chain. In this rule,
the filtering specification is a predicate which always evaluates to
\emph{True} and the action is one of \emph{ACCEPT} or \emph{DROP}
values.

The \emph{SET-MARK} action will be mapped to \emph{SET} instructions,
setting the variable with number 1 to \textit{mark} value.  Thus, for
example: \emph{SET-MARK 12} becomes \emph{\$1=12}.

The \emph{MARK} filtering specification will be mapped to the
\emph{VAR} intermediate filtering specification, checking if the
variable 1 value matches a given constant. So, for example \emph{MARK
  5/0x0F} becomes \emph{\$1 = 5 \& 0x0F}.

\subsubsection{PF}
\label{subsec:pfabstr}

All rules are assigned non-overlapping ranges of \textit{labels}.
Rules within each ruleset have sequential labels.

Since the default action for a ruleset is \emph{PASS} in PF, an
additional rule will be added at the end of each ruleset. In this
rule, the filtering specification is a predicate which always
evaluates to \emph{True} and the target is \emph{ACCEPT}.

Both \emph{ACCEPT} and \emph{DROP} where the \emph{quick} option is present are
mapped to corresponding intermediate language instructions. 

To simulate rules without the \emph{quick} option, a special variable
with number 0 will be used. The rules without the \emph{quick}
option will save their action in this variable. At the end of the
policy, two rules will be added to check if this variable holds a
\emph{DROP} action and, if it does, to trigger it. We do not need to
check for an \emph{ACCEPT} value because it is a default.

For example:

\begin{verbatim}
{filtering-spec} PASS
\end{verbatim}

will be mapped to:

\begin{verbatim}
10 if {filtering-spec} then $0 = 'ACCEPT' ;
...

65534 if $0 = 'DROP' then drop ;
65535 if true then accept ;
\end{verbatim}

\textit{tags} will be stored in a special variable with number
1. Tagging instructions will be translated to:

\begin{verbatim}
if ... then $1 = tagname
\end{verbatim}

Checks for a tag presence will translate to:

\begin{verbatim}
if $1 = tagname then ...
\end{verbatim}

The \textit{inline anchors} definition will be moved from the ruleset they
were defined in to a separate block of rules, terminated with an
unconditional \emph{RETURN} target. In place of their definition,
an unconditional \emph{CALL} statement will be inserted.

Anchor invocation (via \emph{anchor} target) will be replaced with a
\emph{CALL} target.

\subsubsection{IPFW}

Rules are mapped, preserving their numbers as labels. A rule with
label 65536 is added which unconditionally invokes the \emph{DROP}
target.

The \emph{SKIPTO} action is mapped to the \emph{JUMP} target.

\emph{TAG tag-number} is mapped to \emph{\$tag-number =
  'TRUE';}. \emph{UNTAG tag-number} is mapped to \emph{\$tag-number =
  NIL}. The checks for a tag presence are mapped to something like
\emph{20 if \$tag-number = NIL then \ldots}.

\textit{Sets} are just a way to organize policy rules and are not
mapped into the intermediate language. We will consider policy resulting
from selecting rules from all sets which are enabled at the
moment. Enabling different sets will produce multiple different
policies, which can be analyzed separately.

\subsubsection{IPFilter}

All rules are first sorted by ascending group number, preserving order
within a group. Labels are assigned according to this new rule order.

\textit{quick} option handling will be done in the same manner as in
PF, described in Section~\ref{subsec:pfabstr}.

Group invocation (\emph{HEAD} action) is mapped into the \emph{CALL}
instruction. Each group uses it's own variable, with a name equal
to the first rule number in the group to store a resulting action. Thus,
a group starting with rule number 10 will store a resulting action in
a variable with the number 10\footnote{Because \emph{GROUP} calls can be
  nested, the group could indirectly call itself. This should not pose
  any problems, since in \emph{ipfilter} there are no dynamic checks
  and the outcome of the group execution is always the same in the
  context of single packet processing}.

For example:

\begin{verbatim}
{filtering-spec} HEAD 10 DROP
\end{verbatim}

will be mapped to:

\begin{verbatim}
100 if {filtering-spec} then $10 = NIL ;
101 if {filtering-spec} then call 10 ;
102 if {filtering-spec} and $10 = 'ACCEPT' then $0 = 'ACCEPT' ;
103 if {filtering-spec} and ! $10='ACCEPT' then $0 = 'DROP' ;
\end{verbatim}

In the example above, the last two lines should really be three:

\begin{verbatim}
102 if {filtering-spec} and $10 = 'ACCEPT' then $0 = 'ACCEPT' ;
103 if {filtering-spec} and $10 = 'DROP' then $0 = 'DROP' ;
104 if {filtering-spec} and $10 = NIL then $0 = 'DROP' ;
\end{verbatim}

But knowing that in this context, the variable 10 could take only three
values: \emph{'ACCEPT'}, \emph{'DROP'} or \emph{NIL} we can replaces
these three rules with the two rule equivalent shown above.

If the \emph{quick} option were specified, our example would look like:

\begin{verbatim}
{filtering-spec} quick HEAD 10 DROP
\end{verbatim}

mapped to:

\begin{verbatim}
100 if {filtering-spec} then $10 = NIL ;
101 if {filtering-spec} then call 10 ;
102 if {filtering-spec} and $10 = 'ACCEPT' then accept ;
103 if {filtering-spec} and ! $10 = 'ACCEPT' the drop ;
\end{verbatim}

The \emph{SKIP} action will be mapped to the \emph{GOTO} target, converting
relative offset into an absolute rule label.

\newpage\section{Firewall Policy Analysis}
\label{sec:analysis}

The goal of this section is to discuss how \textit{static code
  analysis} could be applied to a firewall policy expressed in
\textit{intermediate language} as introduced in
Section~\ref{sec:abstract-lang}. While our immediate goal is
\textit{anomaly detection}, these techniques could be extended further
for other purposes, such as optimization.

The main premise of our work is that firewall policy is essentially a
program in an imperative programming language, which is executed by
a firewall. The input of the program is an IP packet (defined by a set of
field values) and the outcome is a decision as to how this packet should be
handled (passed through or dropped).

The techniques for analyzing relations between individual rules are
well understood. Some of them are mentioned in
Section~\ref{sec:formalmodel}.  The main challenge is dealing with
control flow constructs, which could impact the order of execution of
individual rules.

The \textit{intermediate language} which we introduced in
Section~\ref{sec:abstract-lang} is highly suitable for the application of
static code analysis. This is a fairly simple imperative programming
language with variables, conditional statements and two simple control
flow constructs: (\emph{JUMP} and \emph{CALL/RETURN}).

Although we cannot assume that any program in this language is
guaranteed to terminate, the programs generated by converting valid
policies from original policy languages will terminate. (due to
restrictions like loop prevention and \emph{GOTO} only pointing to the beginning of
the chain in \textit{Netfilter}, and forward-only \emph{SKIPTO} in
\textit{IPFW}.)

\subsection{Minimal Combining Set Of Intervals}

In this section we will first introduce a notion of a \textit{Minimal
  Combining Set Of Intervals (MCSI)}.  MCSI will be used later on in
policy analysis to represent static checks as an operation on boolean
variables instead of an operation on intervals. This substituion will allows
us to more easily apply Datalog and BDD for policy analysis.

\subsubsection{Definition}

Let $O$ be an arbitrary \textit{set} of \textit{non-empty}
intervals. No assumptions are made about intervals in this set: they
may overlap, be \textit{open}, \textit{closed}, \textit{half-open},
\textit{degenerate}, \textit{bounded}, \textit{unbounded} or
\textit{half-bounded}\cite{mwinterval}. What is described below
applies to \textit{real} or \textit{integer} intervals, or more
generally to intervals defined on totally ordered sets.

We will use upper case letters for \textit{sets}, letters with bar on
top for \textit{intervals} and lower case letters for set members.

We call $Z$ a \textit{Minimal Combining Set (MCSI)} of $O$ iff 1-4 are
all true:

\paragraph{Same Coverage.} Every point which belongs to any of the
intervals in $O$ also belongs to some interval in \textit{MCSI} and
vice versa. More formally:

\begin{eqnarray}
\label{cov1}
\forall \bar{S} \in O, \forall x \in \bar{S}, \exists \bar{A} \in Z, x \in \bar{A} \\
\label{cov2}
\forall \bar{A} \in Z, \forall x \in \bar{A}, \exists \bar{S} \in O, x \in \bar{S}
\end{eqnarray}

\paragraph{Disjoint.} Intervals in \textit{MCSI} are disjoint
(non-overlapping):

\begin{eqnarray}
\label{dj}
\forall \bar{M},\bar{N} \in Z, \neg (\exists x \in \bar{M}, \exists y \in \bar{N}, x=y \land \bar{M} \neq \bar{N})
\end{eqnarray}

\paragraph{Composite.} Any interval in the original set could be
exactly represented by one or more intervals from \textit{MCSI}:

\begin{eqnarray}
\label{comp}
\forall \bar{S} \in O, \exists A_s \in \mathcal{P}(Z),
\biggl[
\begin{split}
(\forall x \in \bar{S}, \exists \bar{A} \in A_s, x \in \bar{A}) \land \\
(\forall \bar{A} \in A_s, \forall x \in \bar{A}, x \in \bar{S})
\end{split}
\biggr]
\end{eqnarray}

\subsubsection{Properties}

Let us prove the \textit{MCSI uniqueness} property which is useful to us:

\begin{theo:mcsiunique}
  There is a single way to represent each interval from the original
  set using intervals from \textit{MCSI}.
\end{theo:mcsiunique}

\begin{proof}
Given the original set $O$ and it's MCSI $Z$, let us assume there is
$\bar{S} \in O$ for which there are two two distinct ways to
represent it using intervals from $Z$. Let us call them $Z^s_1,Z^s_2
\subseteq Z$. Being non-equivalent:

\begin{eqnarray}
Z^s_1 \not\equiv Z^s_1 \leftrightarrow \\
(\exists \bar{A} \in Z^s_1, \bar{A} \notin Z^s_2) 
\lor 
(\exists \bar{A} \in Z^s_2, \bar{A} \notin Z^s_1) \leftrightarrow \\
\label{zneq}
(\exists \bar{M} \in Z^s_1, \forall \bar{N} \in Z^s_2, \bar{M} \neq \bar{N}) 
\lor 
(\exists \bar{N} \in Z^s_2, \forall \bar{M} \in Z^s_1, \bar{N} \neq \bar{M})
\end{eqnarray}

Let as consider $\bar{M} \in Z^s_1$ and $x \in \bar{M}$. According to \eqref{cov2}: $\exists \bar{S} \in O, x \in \bar{S}$.

Now let us consider $Z^s_2$. For $x \in \bar{S}$ according to
\eqref{cov1} $\exists \bar{N} \in Z^s_2, x \in \bar{N}$.

This gives us the following system of equations:

\begin{eqnarray*}
\exists \bar{M} \in Z^s_1, x \in \bar{M} \\
\exists \bar{N} \in Z^s_2, x \in \bar{N} 
\end{eqnarray*}

According to \eqref{dj} this means that $\bar{M} = \bar{N}$. Thus \eqref{zneq} will be always false.
\end{proof}

\subsubsection{Algorithm}

An implementation of the \textit{MCSI} calculation algorithm in Haskell
programming language is presented in Appendix~\ref{mcsisrc}. Below
is an informal description of the algorithm.

For all intervals in the original set, build a combined set of their
\textit{boundaries}. The \textit{interval boundary} is a 3-tuple which
consists of:

\begin{enumerate}
\item Endpoint
\item Boolean flag indicating whether the endpoint is included in the
  interval or not
\item Boolean flag, which is \emph{False} if this endpoint represents
  lower bound or \emph{False} if it represents an upper bound of an
  interval
\end{enumerate}

For example, the interval $[4-10)$ has two boundaries:
$(4,true,true)$ and $(10,false,false)$.

The algorithm is a simple iterative algorithm. It takes a list of
intervals and calculates a list of all their boundaries. Then it proceeds
by \textit{splitting} each interval by each of its boundaries. If no
splits have been performed, the list of intervals is \textit{MCSI}. If
splits have been performed, split intervals are replaced with the results
of the \emph{split} operation and the algorithm repeats from the
beginning.

A \textit{split} operation takes an interval and a \textit{boundary}
(3-tuple). If the endpoint is included, the interval is split into two
non-overlapping sub-intervals by this point. Only one of them would
include the point. The decision which one depends on is whether it was an
opening or closing boundary. Example: Splitting $[0-3]$ by
$(2,true,true)$ will produce $[0-2),[2-3]$. Splitting it by
$(2,true,false)$ will produce $[0-2],(2-3]$.

If the boundary endpoint is excluded, the interval will split into
three non-overlapping sub-intervals: from the interval's lower bound to
the boundary point (not including it), a degenerate interval which
contains the boundary point alone, and finally a third interval, which
includes the bound point to the upper bound of an interval. The
reason for such a split is to exclude the bound point when
combining these sub-intervals to represent some of the original
intervals. Example: splitting $[0-3]$ by $(2,false,true)$ will produce
$[0-2),[2-2],(2-3]$.

We are trying to produce a \textit{minimal} set of intervals. If this
restriction is lifted, it is sufficient to split always into three
sub-intervals, as described in a previous paragraph.

\subsection{Applying MCSI to Static Checks}
\label{subsec:applyingmcsi}

Most firewall rules perform checks on the fields of an IP packet.
There is a limited set of these fields, such as the \textit{source address},
\textit{destination address}, \textit{port}, etc. They all have
numeric values which fall into well-defined intervals. For example, an IP
(version 4) address that could be represented as a 32-bit integer must be in interval
$[0, 2^{32}-1]$. TCP port numbers are in interval $[0, 2^{16}-1]$. These
intervals are always totally ordered sets of individual, discrete
values.

All \textit{static checks} are operating on an individual value from
or on sub-intervals of these intervals. For example, one may check if
a port number is in interval $[0,1023]$ (a privileged port) or if an
IP address belongs to subnet $192.168.1.0/24$ which could be also
converted to an interval $[192.168.1.1, 192.168.1.255]$.

In a firewall policy, we treat the values of each field as belonging to a
distinct domain. For example, there is a domain of source IP addresses
and there is a domain of source TCP port numbers\footnote{It should be
  noted that although for example source port numbers and destination
  port numbers have the same physical type and range they belong to two
  distinct domains}. In our intermediate language, all value checks on
these intervals are static. In other words, IP packet fields are only
checked whenever they belong to \textit{constant intervals},
hard-coded in a policy. This allows us to find all possible constant
intervals used for each domain. Then, for each domain we can calculate
MCSI for this domain.

For example, if we have a domain of TCP port numbers $[0,65535]$ and
the following checks in a policy:

\begin{eqnarray*}
rule1 \leftarrow port = 2 \\
rule2 \leftarrow (port \geq 2) \land (port \leq 3) \\
\end{eqnarray*}

This allows us to identify the following set intervals for domain of TCP
ports: $[2,2],[2,65535],[0,3]$. MCSI representation of this set is:
$[[0-2),[2-2],(2-3),[3-3],(3-65535]]$. Because TCP port numbers are
integers $(2,3)$ interval is equivalent to an empty interval and could
be omitted.

Thus any TCP port in this policy belongs to one of the intervals:

\begin{eqnarray*}
i1 = [1,2) \\
i2 = [2,2] \\
i3 = [3,3] \\
i4 = (3,65535] \\
\end{eqnarray*}

Thus, the policy checks could be expressed as:

\begin{eqnarray*}
rule1 \leftarrow port \in i2 \\
\\
rule2 \leftarrow ((port \in i2) \lor (port \in i3) \lor \\ (port \in i4)) \land ((port \in i3) \lor (port \in i2) \lor (port \in i1)) \\
\end{eqnarray*}

This allows us to represent all static check expressions as a set of
predicates, one per interval from \textit{MCSI}. From this point on we
can forget actual interval values, and treat them as Boolean
variables, assigned result of evaluation of packet fields towards
these intervals. There will be a finite set of these variables.

How could this help us to solve the problem we are tackling? Data and
Control flow analysis relies heavily on the \textit{Control Flow
  Graph}. This is a graph which represents how control can pass
between labels of a program. We can build a control flow graph where
each node will have a list of static checks associated with it. Since
there is a finite number of these checks, if the same condition is
checked in multiple nodes of the graph, they will have exact same
checks. This allows us to represent the control flow graph as a
\textit{Binary Decision Diagram} which is well suitable for analysis
using various algorithms.

\subsection{Unreachable Code Detection}
\label{unreach_theory}

\textit{Unreachable Code Detection} is a kind of \textit{Control Flow
  Analysis}.  

\textit{Unreachable Code} is defined as:
\begin{quotation}
  ``A code fragment is unreachable if there is no control flow path to
  it from the rest of the program. Code that is unreachable can never
  be executed, and can therefore be eliminated without affecting the
  behavior of the program.''\cite{349233}
\end{quotation}

We start by building a control flow graph of a firewall policy. Labels
act as the nodes of this graph. Edges can be associated with a list of
constraints: a static checks which packets must be satisfied for control
flow to traverse this edge. The label is reachable if there exists at least
one path from one of initial labels to this label. The constraints
along the path must not be inconsistent. I.e. it must be possible to
have a packet which satisfies all of them.

\subsection{Live Variable Analysis}
\label{live_theory}

\textit{Live Variable Analysis} is a type of \textit{data flow analysis}
which could be used to find and eliminate \textit{dead code}. In
particular, the code which assigns variables that are always re-assigned
later. In other words, rules which are redundant and could be omitted
from a policy. More formally:

\begin{quotation}
  ``A variable is \textit{live} at the exit from a label, if there exits
  a path from the label to a use of the variable that does not
  re-define the variable.''\cite{nielson1999ppa}
\end{quotation}

Performing live variable analysis using a monotone framework involves
finding a fixed point for a given lattice of finite height and functions
$f$. There are several algorithms to find the fixpoint. For example
Nielson et al. present\cite{nielson1999ppa} \textit{Chaotic Iteration
  algorithm}, \textit{Maximal Fixed Point solution}, and \textit{Meet Over
  all Paths solution}.

In case of multi-trigger policies, when a policy decision is stored in
a (special) variable, we can apply live variable analysis to find all
places where variable assignments are always overwritten
afterwards. Such places indicate \textit{rule shadowing}.

Example:

Sample policy in a native language (\emph{pf}):

\begin{verbatim}
block in on en0 from 192.168.1.10/32 to any 
pass out on en0 from 192.168.1.0/24 to any 
\end{verbatim}

Assuming multi-trigger action, the first rule is always shadowed by the
second one.  

Let us convert this into an intermediate language.

\begin{verbatim}
1 if saddr in 192.168.1.10/32 then 
     $0='drop';
2 if saddr in 192.168.1.00/24 then 
     $0='accept';
# eplilogue
1000 if $0='accept' then 
     accept;
1001 if $0='drop' then 
     drop;
\end{verbatim}

The variable \emph{\$0} is not \textit{live} at the exit from label
\emph{1}. Label \emph{1} corresponds to the first rule in the original
policy, meaning that this rule is redundant and could be omitted.

\newpage\section{Implementation}
\label{sec:implementaion}

We have implemented a simple analyzer, analyzing a program in an
\textit{Intermediate Rule Language} and detecting some potential
anomalies.

The implementation consists of the following modules:

\begin{enumerate}
\item \textit{Parser} is responsible for parsing a program in
  Intermediate Rule Language and representing it as a
  \textit{parse tree}.

\item \textit{Data Flow and CFG Extractor} is responsible for taking
  parsed policy and extracting from it some facts, which will be used in
  further analysis.

\item \textit{Static Analyzer} working with facts produced by previous
  module implements \textit{live variable analysis} and
  \textit{unreachable code detection}.
\end{enumerate}

The first two modules are implemented in Haskell. Haskell is an advanced
purely functional programming language. 

The \textit{Datalog} language is commonly used as an implementation
language for program analysis algorithms\cite{ullman1988pda},
\cite{whaley2005udb}. Datalog is a query language based on the logic
programming paradigm. It is a subset of the \textit{Prolog} logic
programming language.

\begin{quotation}
  ``Analyses expressed in a few lines of Datalog can take hundreds to
  thousands of lines of code in a traditional
  language.''\cite{whaley2005udb}
\end{quotation}

We use Datalog for \textit{Static Analyzer Module} implementation.

Source code for all modules (excluding unit tests and build files) is
included in Appendix~\ref{src}. We will discuss each of these modules in
more detail below.

\subsection{Parser Module}

This module is parsing concrete syntax of our \textit{Intermediate
  Policy Language} as defined in
Section~\ref{subsec:abstract-lang-concrete-syntax} into a
\textit{Parse Tree}. Parse tree structure is pretty much
defined by \textit{abstract syntax} as defined in
Section~\ref{subsec:abstract-lang-abstract-syntax}. We are using
Haskell data structures to represent it.

Coding a parser by hand is a laborious task, and tools called
\textit{parser generators} are commonly used to automate it. These
tools usually accept high-level definition of syntax of the input language
and generate parser code in the target language. The tool we have chosen
for this purpose is called \textit{Happy}, a monadic parser generator
for Haskell\cite{gill1995hpg}.

Parser grammar specification in annotated BNF syntax for
\textit{Happy} is included in Appendix~\ref{parsersrc}.

The type of parser we have implemented is called a \textit{monadic
  parser}.  We are using a variant of \textit{Exception Monad} (also
known as \textit{Error Monad}) for error handling. The monad we use is
defined by type constructor \emph{P}, bind operation \emph{thenP} and
return operation \emph{returnP}.

We define tokens used in our language using the \emph{\%token} directive.
The lexical analyzer, responsible for splitting a source file into
tokens is pretty trivial and hand-coded in Haskell, and its source is
included in \emph{PolicyLang.y} as a \emph{lexer} function. The lexer
is also monadic. It is called by the parser to emit new tokens, and is
passed a \textit{continuation} as an argument. The new token is read and a
continuation is called with it.

The bulk of the grammar consists of \textit{production rules}. Each
rule consists of a \textit{non-terminal symbol} on the left side
followed by one or more \textit{expansions} on the right
side. Expansions have Haskell code associated with them (in curly
brackets) which specify the \textit{value} of each expansion. The
parser matches a stream of tokens produced by lexer towards productions,
and builds a \textit{Parse Tree} from values emitted by
productions.

The data types for the parse tree as defined in separate Haskell modules:
\textit{NetworkData} and \textit{Policy}, the source of which is also
included in Appendix~\ref{parsersrc}.

The \textit{NetworkData} Haskell module contains definitions of data types
for basic network concepts such as \textit{IP address}\footnote{In the current implementation, we work only with IP version 4 addresses although our algorithms could be extended to work with IP version 6 addresses as well}, \textit{IP
  Network}, and \textit{Netmask} as well as utility functions for working
with them, such as converting between \textit{CIDR} and
\textit{netmask} notations. It also contain instances of the
\emph{Interval} class for IP addresses and TCP port numbers.

Haskell module \textit{Policy} contains definitions of data types for
firewall policies in intermediate language: types such as
\emph{PolicyRule}, \emph{StaticCheck}, and \emph{Target}.

\subsection{Data Flow and CFG Extractor Module} 
\label{subsection:cfgext}

The purpose of this module is for a given program (parser module output) to
produce a set of facts for the Datalog analyzer.

The facts it produces are:

\begin{enumerate}
\item List of internal and external labels
\item List of initial labels
\item List of final labels
\item List of variables
\item At what labels writes to what variables occur
\item At what labels reads from what variables occur
\end{enumerate}

Additionally, it generates Datalog predicates for the control flow
graph. Because of the predicate structure we have chosen for our Datalog
analysis (see next section), it also generates analysis predicates
with appropriate arity. The arity depends on policy being analyzed.

Let us first look at the notion of \textit{internal} vs \textit{external}
labels. In our intermediate policy language all rules have
labels. This is what in policy analysis we will refer to as
\textit{external} labels. Static code analysis also operates
with \textit{labels}. However these labels (which we will
call \textit{internal labels}) are more granular, since we
should be able to address individual parts of the rule. 
Each \textit{external label} is corresponding to a rule which
hse the following general structure:

\begin{verbatim}
label 'if' filteringspec 'then' targetspec ';'
\end{verbatim}

To reflect the fact that control flow might now reach \emph{target}
spec if \emph{filteringspec} is not matched, we need to distinguish at
least two \textit{internal labels} here: one for filtering spec and
one for target spec. Let us call them $l_{if}$ and $l_{then}$
and associate them with filtering specification and target specification
respectively. After the control flow reaches $l_{if}$, if filtering
specification is satisfied it would proceed to $l_{then}$. Otherwise it
would proceed to the next rule. 

\begin{figure}[htp]
\centering
\includegraphics[width=3in]{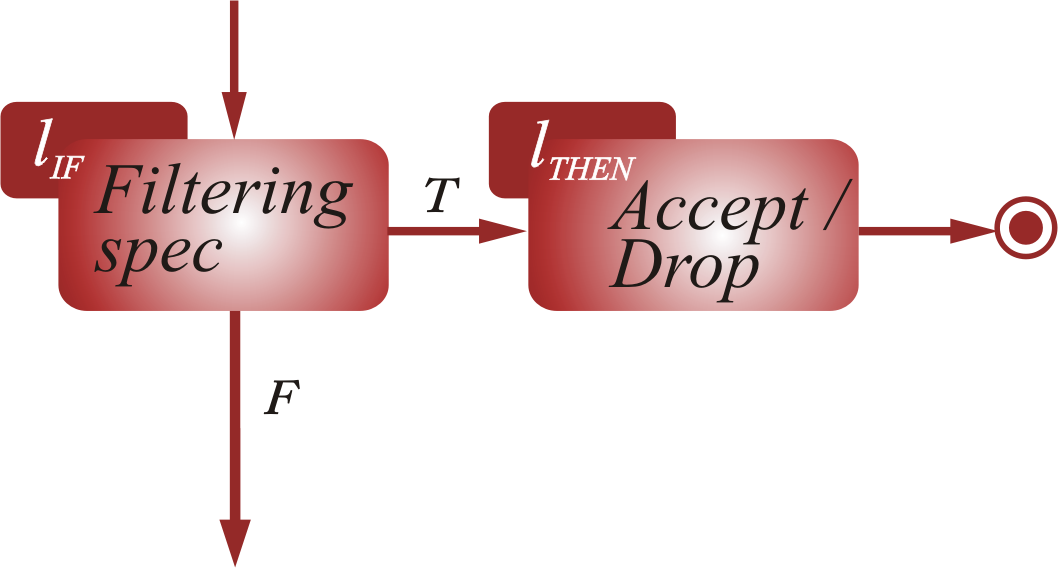}
\caption{Control flow graph of of policy statement with \emph{ACCEPT/DROP} target}
\label{fig:accept_flow}
\end{figure}

\textit{Target specification} in our intermediate language can have
only 6 possible values:

If it is either \emph{ACCEPT} or \emph{DROP} then $l_{then}$ is
final and control will not proceed past it. The control flow graph
for this case is shown on Figure~\ref{fig:accept_flow}.

If it is \emph{JUMP}, it would proceed to the label specified as a
parameter of \emph{JUMP}. To be more precise, it will proceed to the
internal label which corresponds to the \textit{if} part of the external label
specified as the \emph{JUMP} argument. So for \emph{JUMP X} the control
flow will proceed to $X_{if}$.

\begin{figure}[htp]
\centering
\includegraphics[width=3in]{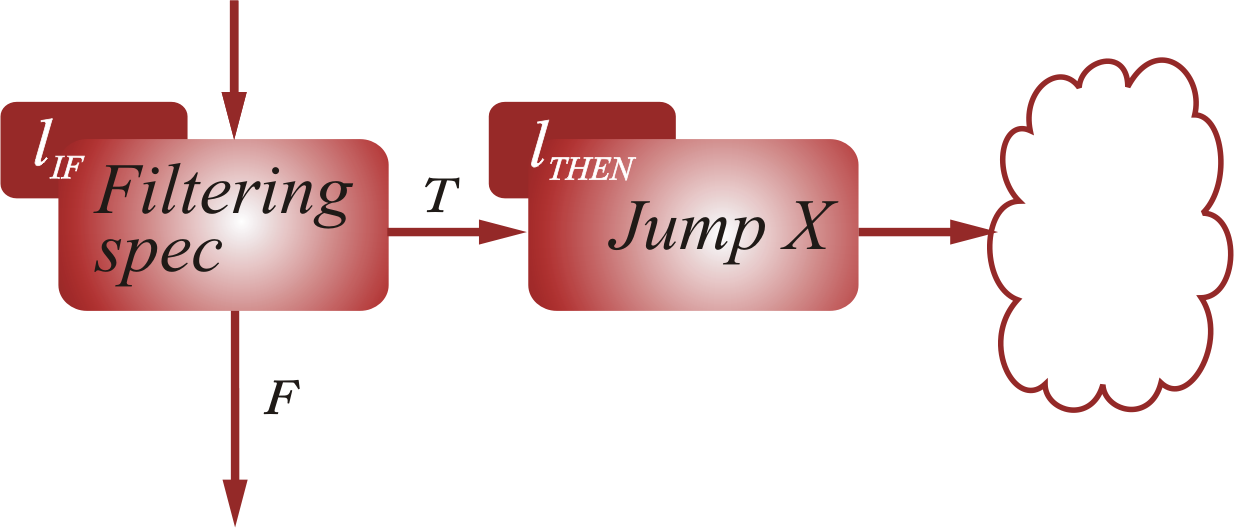}
\caption{Control flow graph of of policy statement with \emph{JUMP} target}
\label{fig:jump_flow}
\end{figure}

The control flow graph for this case is shown on Figure~\ref{fig:jump_flow}.

If it is \emph{SET}, the control will always proceed to the next rule
directly if the static check condition is not satisfied, and via the
$L_{then}$ label otherwise.

\begin{figure}[htp]
\centering
\includegraphics[width=2.2in]{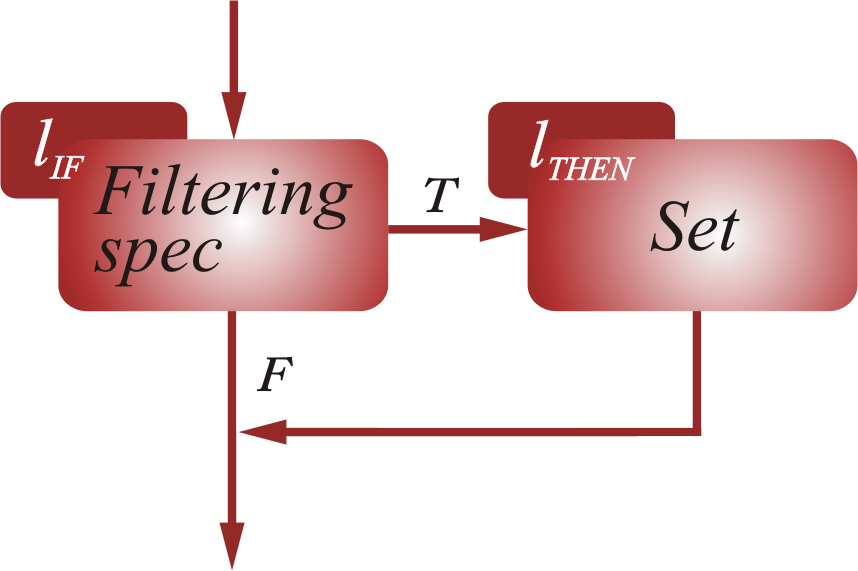}
\caption{Control flow graph of of policy statement with \emph{SET} target}
\label{fig:set_flow}
\end{figure}

The control flow graph for this case is shown on Figure~\ref{fig:set_flow}.

\begin{figure}[htp]
\centering
\includegraphics[width=3in]{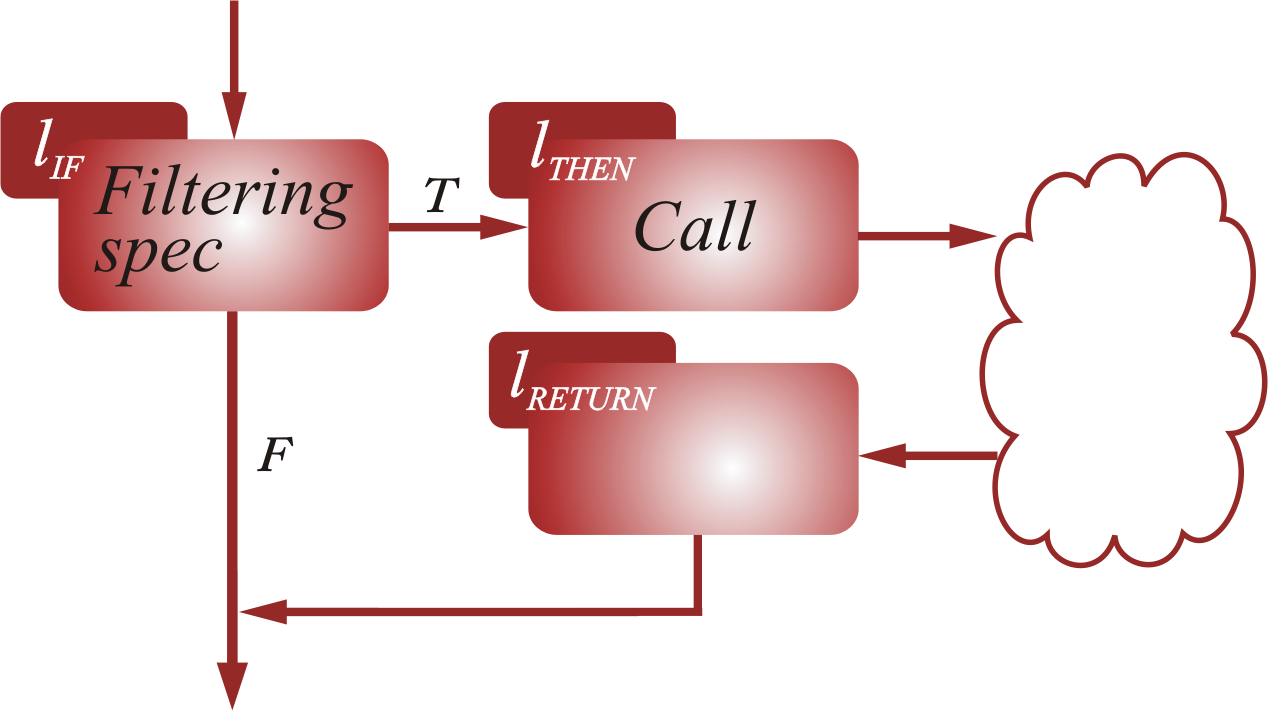}
\caption{Control flow of of policy statement with \emph{CALL} target}
\label{fig:call_flow}
\end{figure}

\begin{figure}[htp]
\centering
\includegraphics[width=3in]{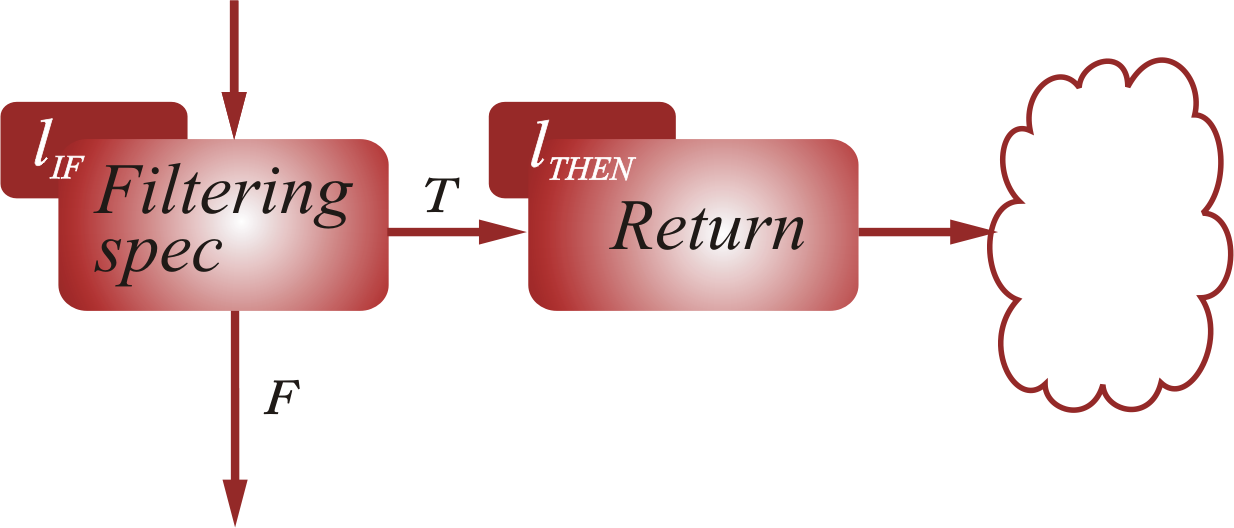}
\caption{Control flow of of policy statement with \emph{RETURN} target}
\label{fig:return_flow}
\end{figure}

Now, we come to a more complex case: \emph{CALL/RETURN}. The control
flow for this case is shown on
Figures~\ref{fig:call_flow}~and~\ref{fig:return_flow}. A \emph{CALL}
target will cause control first to flow to the label specified as the
\emph{CALL} argument, but it could eventually return from it. To
describe this we define another internal label $l_{return}$ which is
where control flow would return after the call. After $l_{return}$
control flow will invariably proceed to the next rule. A \emph{RETURN}
target could cause control flow to come back to the appropriate
$l_{return}$. Determination of which one is a complex problem which is
part of \textit{interprocedural analysis}. For imperative,
non-functional languages it is hard to solve and one common approach
is to make a conservative assumption that any \emph{RETURN} target
could potentially return flow control to any \emph{CALL} statement in
the program. This so-called conservative analysis allows us to perform
some practical dataflow analysis, which might not be exhaustive (will
not detect all possible cases), but will still be useful. For the
purpose of this prototype we decided to ignore \emph{CALL} and
\emph{RETURN} targets, treating them as empty (doing nothing). Adding
support for them could be a direction for future work.

So each \textit{original} label could be mapped into two (or three if the
target is \emph{CALL}) \textit{internal labels}).  Internal labels will
be used in static code analysis. The analysis will produce some
conclusions about internal labels (for instance, that it is
unreachable). Based on these conclusions we can try to infer
conclusions about policy rules referenced by original labels. For
example if all \textit{internal} labels corresponding to an original
label are unreachable we can conclude that the policy rule with this
original label is also unreachable and can be safely removed from
the policy.

\textit{Initial Labels} are entry points of the program. In our
intermediate language, policy processing always starts with the rule which
has the external label with the lowest number. Thus there would be only
one initial label. Since for code analysis we operate in internal
labels, it would be an $l_{if}$ internal label corresponding to the external
label with the lowest number.

\textit{Final Labels} are labels at which program execution
stops. According to our definition of an intermediate language,
rules processing stops when when the \emph{ACCEPT} or \emph{DROP}
target specification is triggered or when control reaches the end of
the policy (last rule). So all $l_{then}$ labels in the rules with an
\emph{ACCEPT} or \emph{DROP} target will be treated as finals, plus
$l_{then}$ label of the last policy rule if it has \emph{SET} as an
action.

\subsection{Static Analyzer Module}

This module is implemented in Datalog. It takes some facts generated
by Data flow and CFG extraction module as described in
Section~\ref{subsection:cfgext} and tries to infer some information
about the original program. Our analysis is targeting the discovery of
``anomalies'': possible inefficiencies or errors in the program. In
particular, we perform two analyses: \textit{Unreachable Code
  Detection} and \textit{Live Variable Analysis}. The theory behind
these analyses are described in Sections \ref{unreach_theory} and
\ref{live_theory} respectively. In this subsection we will discuss
mostly implementation aspects.

As mentioned earlier, we are using the Datalog language to implement
this module. The implementation we use is
\textit{bddbddb}\cite{whaley:bbd}. It is developed at Stanford
University, and written in the Java programming language.

\subsubsection{Generated Facts}

A Datalog program consists of \textit{facts} and \textit{rules}. Both
facts and rules are expressed in the form of \textit{Horn clauses} in
form: $ L_0 \leftarrow L_1, \ldots, L_n$ where $L_i$ is predicate in form
$p(t_1, \ldots, t_n)$ in which $t_n$ are \textit{terms}. Terms could be
\textit{constants} or \textit{variables}.

In \textit{bddbddb} all terms are mapped to integer values in their respective \textit{domains}.

The domains we define are:

\begin{enumerate}
\item $O$ -- a domain of \textit{original} labels
\item $L$ -- a domain of \textit{internal} labels
\item $V$ -- a domain of variables
\item $B$ -- a special domain of Boolean constants
\end{enumerate}

All domains in Datalog have size:

\begin{quotation}
``A domain $D \in \mathcal{D}$ has size $size(D) \in \mathbb{N}$, where $\mathbb{N}$ is the set of natural numbers. We require that all domains have finite size. The elements of a domain $D$ is the set of natural numbers $0 \ldots size(D)-1$.''\cite{whaley2005udb}
\end{quotation}

The domain declarations are generated with calculated sizes. To
minimize domain size, values in domains $L$ and $V$ are generated to be
sequential, without gaps.

Sample domains declaration we generate:

\begin{verbatim}
### Domains
O 1021 olabels.map
L 11 labels.map
V 1000 vars.map
B 2
\end{verbatim}

For each domain we specify size and location of the \textit{EDB} file. In
Datalog, facts can be stored in external relational storage, called
\textit{Extension Database (EDB)} or defined in a Datalog program,
referred as \textit{Intensional Database (IDB)}. To simplify our
implementation and so as not to deal with IDB data formats we only use IDB
for our facts. So, although we specify EDB file names for domains, they
are not used.

The facts we operate with are:

\begin{enumerate}

\item \emph{olabel} predicate defines the relationship between \textit{internal} and
  \textit{external} labels. This fact, strictly speaking, is not
  needed for analysis and is used mostly to provide more user-friendly
  reporting, using \textit{external} label numbers.

\item \emph{label} predicate defines all known internal labels. (Since
  these are the labels we work with, we will refer to them just as
  ``labels'' from now on).

\item \emph{init} predicate defines entry points (labels)

\item \emph{final} predicate defines exit points (labels)

\item \emph{var} predicate defines dynamic variables used in program.

\item \emph{write} predicate defined on variable number and label pairs, states that at given label value of given variable is changed (written to).

\item \emph{read} predicate defined on variable number and label pairs, states that at given label the value of given variable is accessed (read from).
\end{enumerate}

Here is an example of some generated facts:

\begin{verbatim}
label(0).
label(1).
label(3).
label(4).
label(6).
label(7).
label(9).
label(10).
olabel(1000,0).
olabel(1000,1).
olabel(1001,3).
olabel(1001,4).
olabel(1010,6).
olabel(1010,7).
olabel(1020,9).
olabel(1020,10).
final(1).
final(10).
var(0).
var(1).
read(3,1).
write(7,0).
init(0).
\end{verbatim}
\label{olablessample}

\subsubsection{Generated Control Flow Predicates}

In addition to facts, we also generate some
Datalog predicates based on source policy, which represent a control flow graph. These
predicates have a variable number of parameters and in their most general form
could be defined as:

\begin{eqnarray*}
succ(& l_0, l_1,\\
& saddrvariables, sportvariables, \\
& daddrvariables, dportvariables, \\
& protocolvariables \\
).
\end{eqnarray*}

These predicates describe how control could pass from $l_0$ to $l_1$.

In most cases control flow is conditional -- it depends on values of
packet fields being inspected and values of dynamic variables. At the
point of control flow definition it is difficult to reason about the
values of dynamic variables, so we will assume the dynamic check, if
present, can be evaluated to both \emph{True} and \emph{False}, and that
both control flow branches are possible. For a static check we could very
well assume that they will evaluate to the same values for all rules in the
policy, since packet fields do not change durring processing. So
applying MCSI algorithm as described in
Section~\ref{subsec:applyingmcsi}, we represent each part of the static
check (source address check, source port check, destination address
check, destination port check, and protocol check) as a group of Boolean
variables. Each variable could evaluate to \emph{True} if the packet field
value belongs to the respective interval or \emph{False} if not. These
variables (which we will collectively call \textit{static check
  variables} could be grouped as: $saddrvariables$, $sportvariables$,
$daddrvariables$, $dportvariables$, $protocolvariables$.

Let us now revisit control flow diagrams at
Figures~\ref{fig:accept_flow},~\ref{fig:jump_flow},~\ref{fig:set_flow},~\ref{fig:call_flow}~and~\ref{fig:return_flow}. There
are two kind of control flow transitions: conditional and
unconditional. Representing unconditional parts of a control flow graph
is very straightforward: the predicate would not depend on values of
static check variables which in Datalog would be defined as
\textit{anonymous variables}, denoted by an underscore character. For
example predicate:

\begin{verbatim}
succ(3,6,_,_,_,_,_,_).
\end{verbatim}

denotes that control flow from label 3 could pass to label 6
regardless of the fields values of the examined packet.

Conditional control flow predicates depend on values of \textit{static
  check variables}. Static check is satisfied if:

\begin{eqnarray*}
staticcheck \leftarrow & \\
& saddrvariables \land sportvariables \land  \\
& daddrvariables \land dportvariables \land \\
& protocolvariables
\end{eqnarray*}

Each static check variable group can contain one or more
variables (per MCSI split). At least one of them must evaluate to
\emph{True}. For example: $saddrvariables= saddr_0 \lor saddr_n$. So
the full form of static check is:

\begin{eqnarray*}
staticcheck (\ldots) \leftarrow & \\
& (saddr_0 \lor saddr_1 \ldots \lor saddr_n) \land  \\
& (sport_0 \lor sport_1 \ldots \lor sport_n) \land  \\
& (daddr_0 \lor daddr_1 \ldots \lor daddr_n) \land  \\
& (dport_0 \lor dport_1 \ldots \lor dport_n) \land  \\
& (proto_0 \lor proto_1 \ldots \lor proto_n)
\end{eqnarray*}

One way to decompose it into \textit{Horn clauses} is:

\begin{eqnarray*}
staticcheck \leftarrow & \\
& \neg none_n(saddr_0, saddr_1 \ldots, saddr_n) \land  \\
& \neg none_n(sport_0, sport_1 \ldots, sport_n) \land  \\
& \neg none_n(daddr_0, daddr_n \ldots, daddr_n) \land  \\
& \neg none_n(dport_0, dport_n \ldots, dport_n) \land  \\
& \neg none_n(proto_0, proto_n \ldots, proto_n)
\end{eqnarray*}

where:

\[
none_n(v_0, v_1, \ldots, v_n) \leftarrow \neg v_0 \land \neg v_1 \ldots \neg v_n
\]

So for each policy rule, we generate one or more $succ$ predicates. For the static check part we generate different flow for cases where static check is satisfied and where it is not. 

For example, for rule:

\begin{verbatim}
1001 if saddr in {[10.10.10.1,10.10.10.10],[20.0.0.0,20.1.1.1]} 
        sport in [1,100] and $999=5 
     then jump 1020;
\end{verbatim}

We will generate following $succ$ predicates:

\begin{verbatim}
# Label 1001
#  unconditional:
succ(4,9,_,saddr1,saddr2,_,_,sport0).
#  if condidion satified
succ(3,4,_,saddr1,saddr2,_,_,sport0) :- !none(0,saddr1,saddr2,0,0),
                                        !none(sport0).
#  if condidion not satified
succ(3,6,_,saddr1,saddr2,_,_,_) :- none(0,saddr1,saddr2,0,0).
succ(3,6,_,_,_,_,_,sport0) :- none(sport0).
\end{verbatim}

The example above uses internal labels. Mapping between internal and external
labels for this example is shown on page~\pageref{olablessample}.

\subsubsection{Unreachable Code Detection}

This is a first, simple analysis of this module implementation. The theory is
discussed in Section~\ref{unreach_theory}. The top level goal for
this analysis is an $unreachable/1$ predicate which is $true$ for labels
which are not reachable. A label is deemed \textit{reachable} if there
is a path from one \textit{initial label} to it. A path (expressed
via $path$ predicate) is defined as follows:

\begin{verbatim}
path(A,B,x1,x2,x3,x4) :- A=B, label(A), label(B).
path(A,C,x1,x2,x3,x4) :- succ(A,B,x1,x2,x3,x4), 
                         path(B,C,x1,x2,x3,x4), 
                         label(A), label(B), label(C).
\end{verbatim}

In this particular example we have four \textit{static check
  variables} (named $x1,x2,x3,x4$) corresponding to whenever input
packet fields belong to various IP address ports or protocol ranges
mentioned in the the policy. The number of static check variables may
vary depending on the policy being analyzed. The $path$ predicates
with required arity are generated by Haskell code.

\subsubsection{Live Variable Analysis}

The second analysis we chose to implement is Live Variable
Analysis. The theory discussed in Section~\ref{live_theory}.

The Datalog definition of this analysis is just a few lines of code:

\begin{verbatim}
dead(L) :- write(L,V), !live(L,V).

live(L,V) :- init(Li), path(Li,L,x1,x2,x3,x4), read(Lr,V), 
             readonlyPath(L,Lr,V,x1,x2,x3,x4).

path(A,B,x1,x2,x3,x4) :- A=B, label(A), label(B).
path(A,C,x1,x2,x3,x4) :- succ(A,B,x1,x2,x3,x4), 
                         path(B,C,x1,x2,x3,x4), 
                         label(A), label(B), label(C).

readonlyPath(A,B,_,x1,x2,x3,x4) :- A=B, label(A), label(B).
readonlyPath(A,B,_,x1,x2,x3,x4) :- succ(A,B,x1,x2,x3,x4), 
                                   label(A), label(B).
readonlyPath(A,C,V,x1,x2,x3,x4) :- readonlyPath(A,B,V,x1,x2,x3,x4),
                                   succ(B,C,x1,x2,x3,x4), 
                                   var(V), label(A), label(B), label(C),
                                   !write(B,V).
\end{verbatim}

In this particular example we have four \textit{static check
  variables} (named $x1,x2,x3,x4$) corresponding to whenever input
packet fields belong to various IP address ports or protocol ranges
mentioned in the the policy. The number of static check variables may
vary depending on the policy being analyzed. The $path$ and
$readonlyPath$ predicates with required arity are generated by Haskell
code.

\begin{figure}[htp]
\centering
\includegraphics[width=4in]{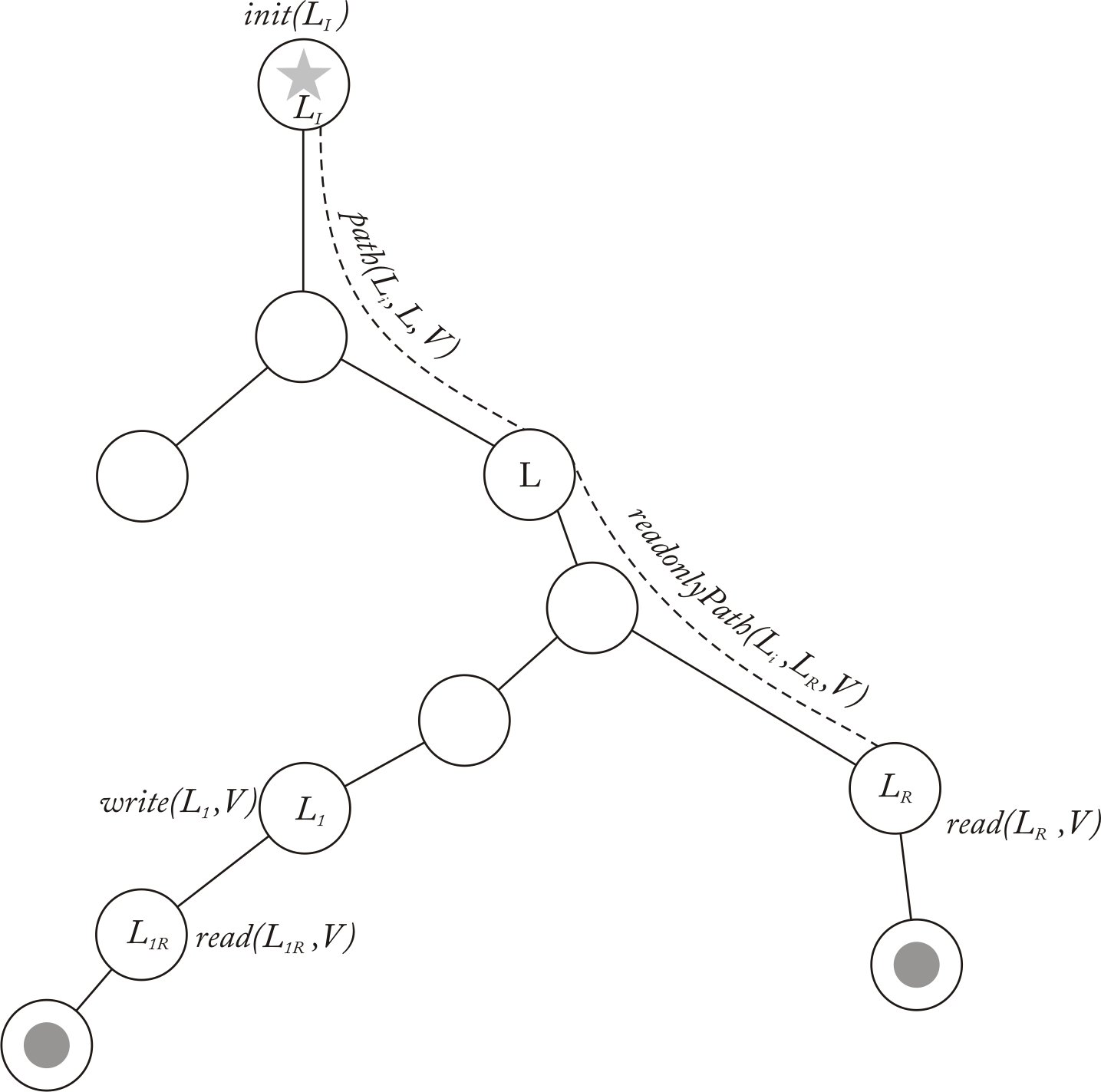}
\caption{Live Variable Analysis example}
\label{fig:live_path}
\end{figure}

The graphical illustration of our implementation of live analysis
can be seen on Figure~\ref{fig:live_path}, which shows a simple control
flow graph. Variable $V$ is \textit{live} at exit from label $L$
because the following holds true:

\begin{enumerate}
\item Variable $V$ is written into at labels $L$ (also at $L_1$)
\item There is a path from one of initial labels ($L_i$) to $L$.
\item There are labels where it is read ($L_R$, $L_{1R}$)
\item There is at least one \textit{read-only path} (path which does not overwrite variable $V$) from $L$ to label where it is read ($L_r$).
\end{enumerate}

In this example, the path from $L$ to $L_{1R}$ is not read-only, because
the variable is overwritten at $L_1$.

Strictly speaking our implementation is little more specific than
``standard'' live variable analysis. In particular we only concern
ourselves with places where variables are written and attempt to
detect non-live variables at the exit point from labels where they are
written. This allows us to detect unnecessary writes.

Additionally, in order to minimize search space, we only analyze
labels which are \textit{reachable}, ignoring \textit{unreachable
  labels}.

Finally, in our analysis we check whether a variable is ever read. We will
report as \textit{dead}, all labels at which variables are written but
never subsequently read from.

\newpage\section{Examples}
\label{sec:examples}

Let us apply our analysis to several simple cases:

\subsection{Example 1}

Original policy in \textit{PF} policy language:

\begin{verbatim}
block in on en0 from 192.168.1.10/32 to any 
pass out on en0 from 192.168.1.0/24 to any 
\end{verbatim}

Since \textit{PF} is using \emph{multi-trigger} actions, we are using
special variable \emph{\$0} to store outcome.

\begin{figure}[htp]
\centering
\includegraphics[width=5in]{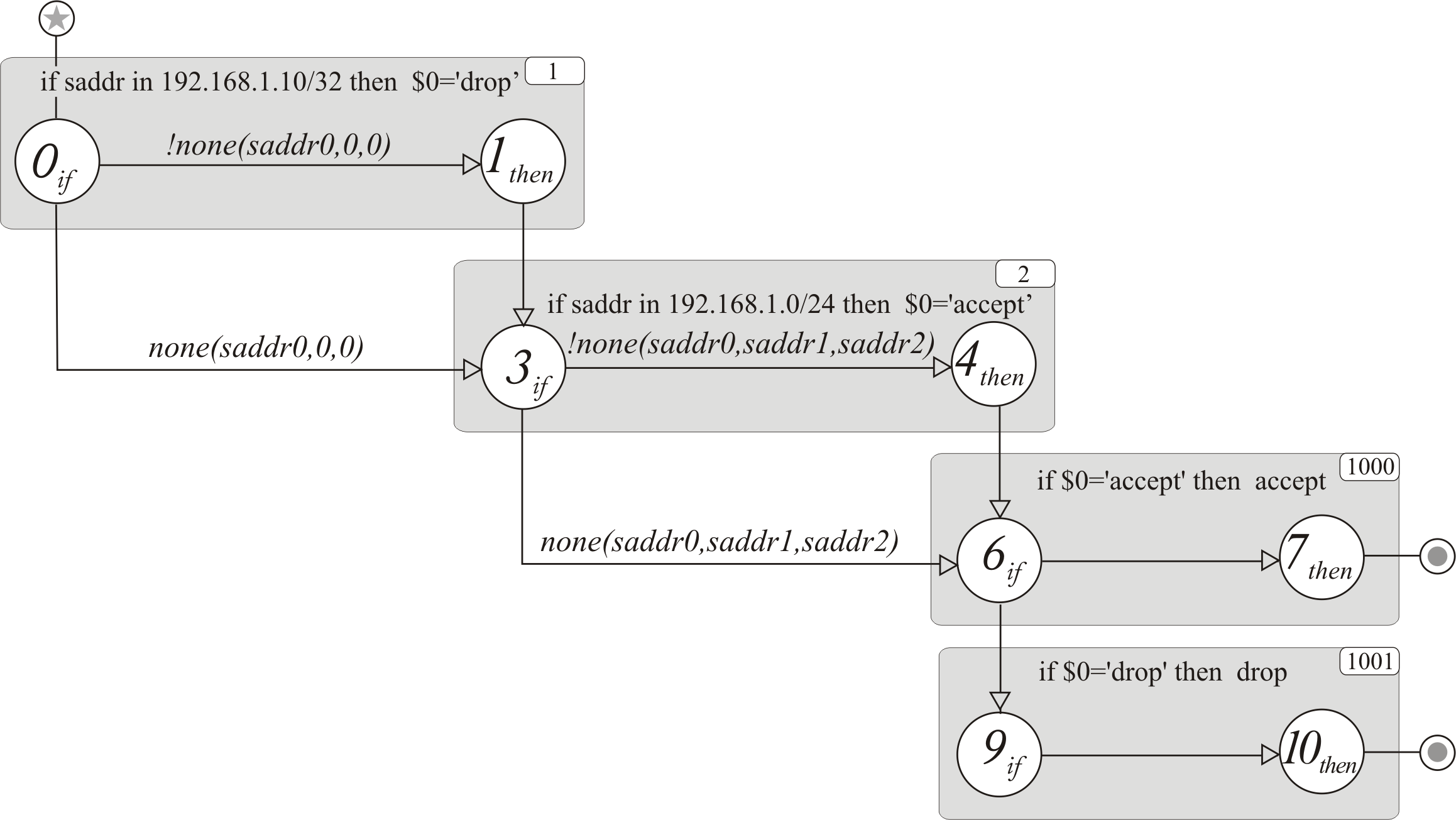}
\caption{Control flow graph for Example 1}
\label{fig:prog_1}
\end{figure}

The same policy, translated to an \textit{Intermediate Policy Language}:

\begin{verbatim}
1 if saddr in 192.168.1.10/32 then 
     $0='drop';
2 if saddr in 192.168.1.0/24 then 
     $0='accept';

# eplilogue
1000 if $0='accept' then 
     accept;
1001 if $0='drop' then 
     drop;
\end{verbatim}

The control flow graph for this program is shown in
Figure~\ref{fig:prog_1}.

Analysis detects that the \$0 assignment in the rule with label 1 is redundant,
since it is always overwritten by the rule with label 2. That means that
that this rule is redundant and could be removed from the policy
without affecting its semantics.

\subsection{Example 2}

Sample policy, already in \textit{Intermediate Policy Language}:

\begin{verbatim}
1000 if saddr in [192.168.1.10,192.168.1.10] 
     then drop;
1001 if saddr in {[10.10.10.1,10.10.10.10],[20.0.0.0,20.1.1.1]} 
        sport in [1,100] and $999=5 
     then jump 1020;
1010 if saddr in [192.168.1.0,192.168.1.255] 
     then $888='1';
1020 if saddr in [192.168.1.0,192.168.1.255] 
     then accept;
\end{verbatim}

\begin{figure}[htp]
\centering
\includegraphics[width=5in]{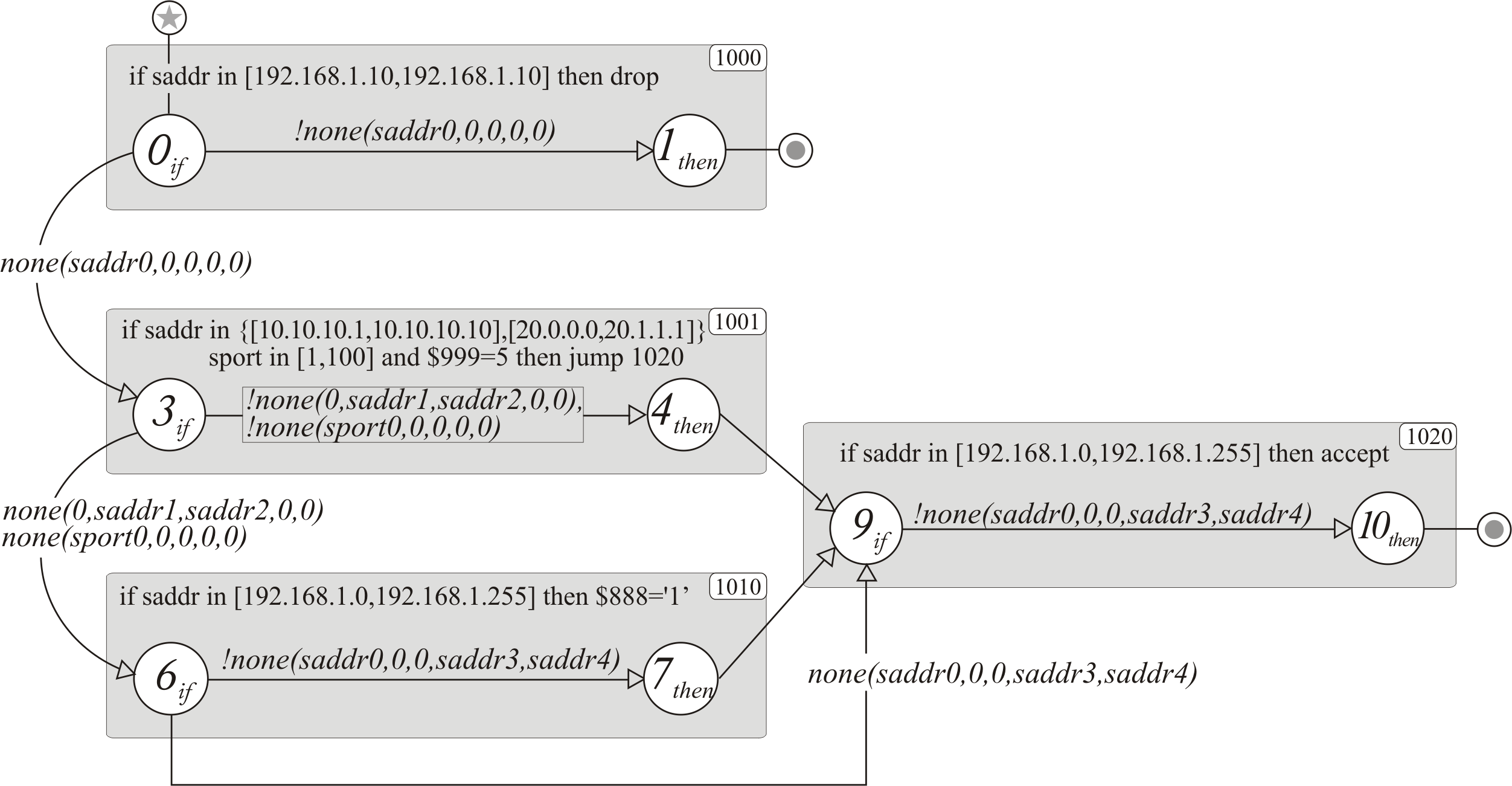}
\caption{Control flow graph for Example 2}
\label{fig:prog_2}
\end{figure}

The control flow graph for this program is shown in
Figure~\ref{fig:prog_2}.

Analysis detects that the \$888 assignment in the rule with label 1010 has no
effect since this variable will never be read afterwards.  That means
that this rule is redundant and can be removed from the policy
without affecting its semantics.

\subsection{Example 3}

\begin{figure}[htp]
\centering
\includegraphics[width=5in]{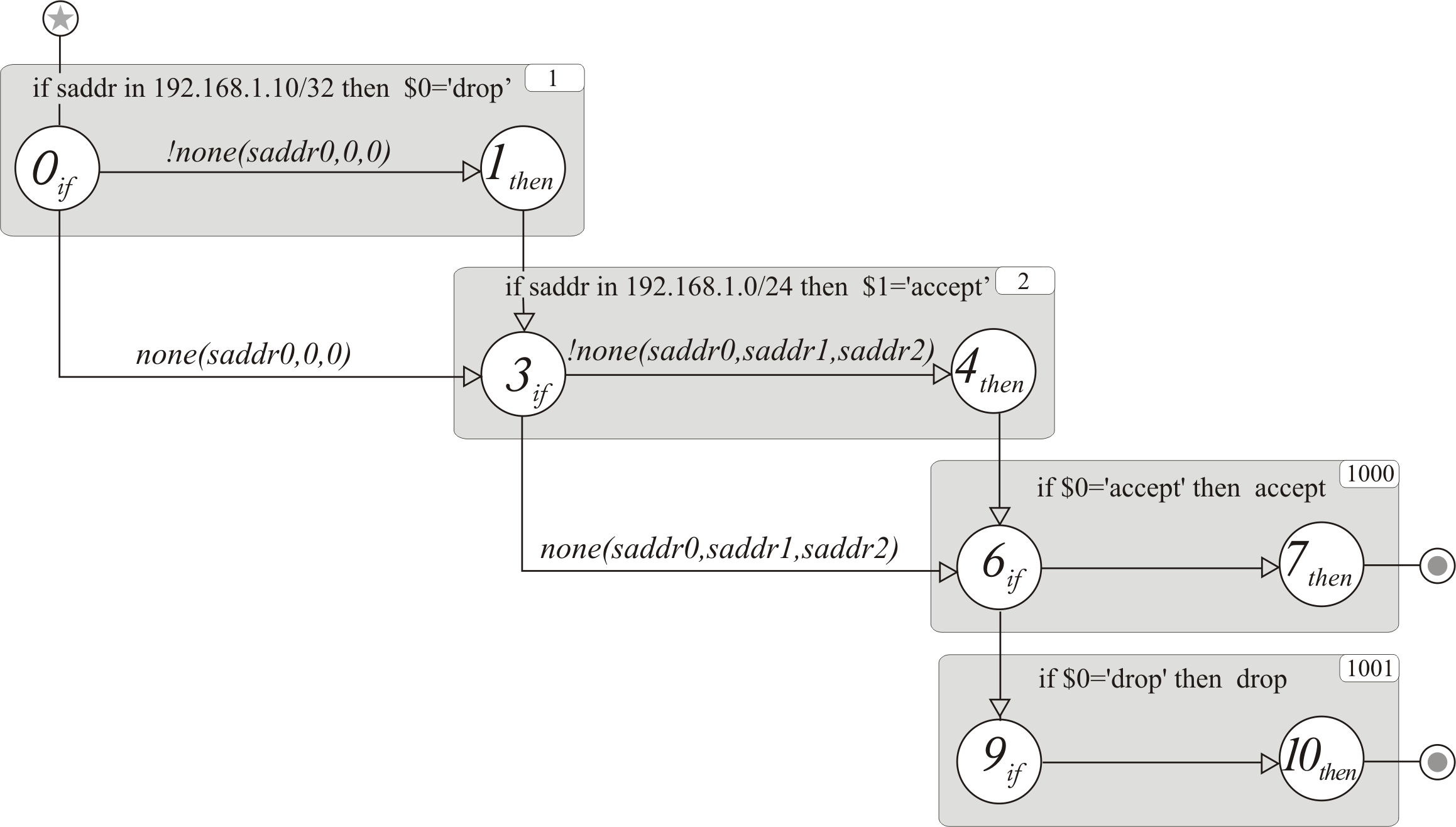}
\caption{Control flow graph for Example 3}
\label{fig:prog_4}
\end{figure}

Sample policy, already in \textit{Intermediate Policy Language}:

\begin{verbatim}
1 if saddr in 192.168.1.10/32 then 
     $0='drop';
2 if saddr in 192.168.1.0/24 then 
     $1='accept';

# eplilogue
1000 if $0='accept' then 
     accept;
1001 if $0='drop' then 
     drop;
\end{verbatim}

The control flow graph for this program is shown in
Figure~\ref{fig:prog_4}.

Analysis detects that \$1 assignment in rule with label 2 has no
effect since this variable will never be read afterwards.  That means
that this rule is redundant and could be removed from the policy
without affecting its semantics.

\subsection{Example 4}

\begin{figure}[htp]
\centering
\includegraphics[width=5in]{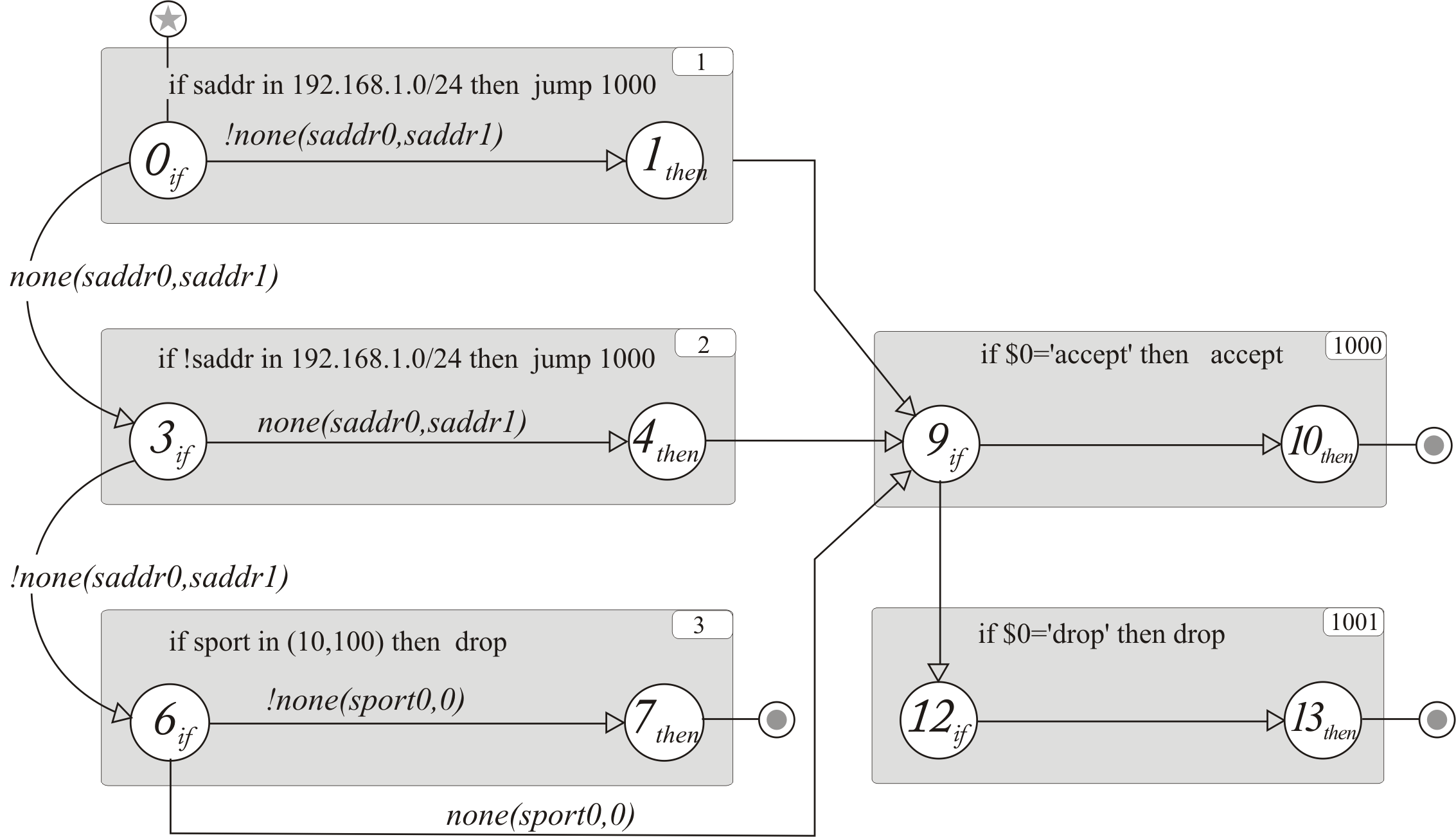}
\caption{Control flow graph for Example 4}
\label{fig:prog_7}
\end{figure}

Sample policy, already in \textit{Intermediate Policy Language}:

\begin{verbatim}
1 if saddr in 192.168.1.0/24 then 
     jump 1000;

2 if !saddr in 192.168.1.0/24 then 
     jump 1000;

3 if sport in (10,100) then 
     drop;

# eplilogue
1000 if $0='accept' then 
     accept;
1001 if $0='drop' then 
     drop;
\end{verbatim}

The control flow graph for this program is shown of
Figure~\ref{fig:prog_7}.

Analysis detects that label 3 is unreachable. Since labels 1 and 2
check for opposite conditions, control flow will
proceed to label 1000 in any case, never reaching label 3.

\subsection{Example 5}

Sample policy, already in \textit{Intermediate Policy Language}:

\begin{verbatim}
1 if saddr in 192.168.1.0/16 then 
     jump 1000;

2 if saddr in {192.168.1.0/24, 192.168.1.0/32, 
              (192.168.1.20,192.168.1.23)} then 
     jump 1000;

3 if sport in (10,100) then 
     drop;

# eplilogue
1000 if $0='accept' then 
     accept;
1001 if $0='drop' then 
     drop;
\end{verbatim}

\begin{figure}[htp]
\centering
\includegraphics[width=5in]{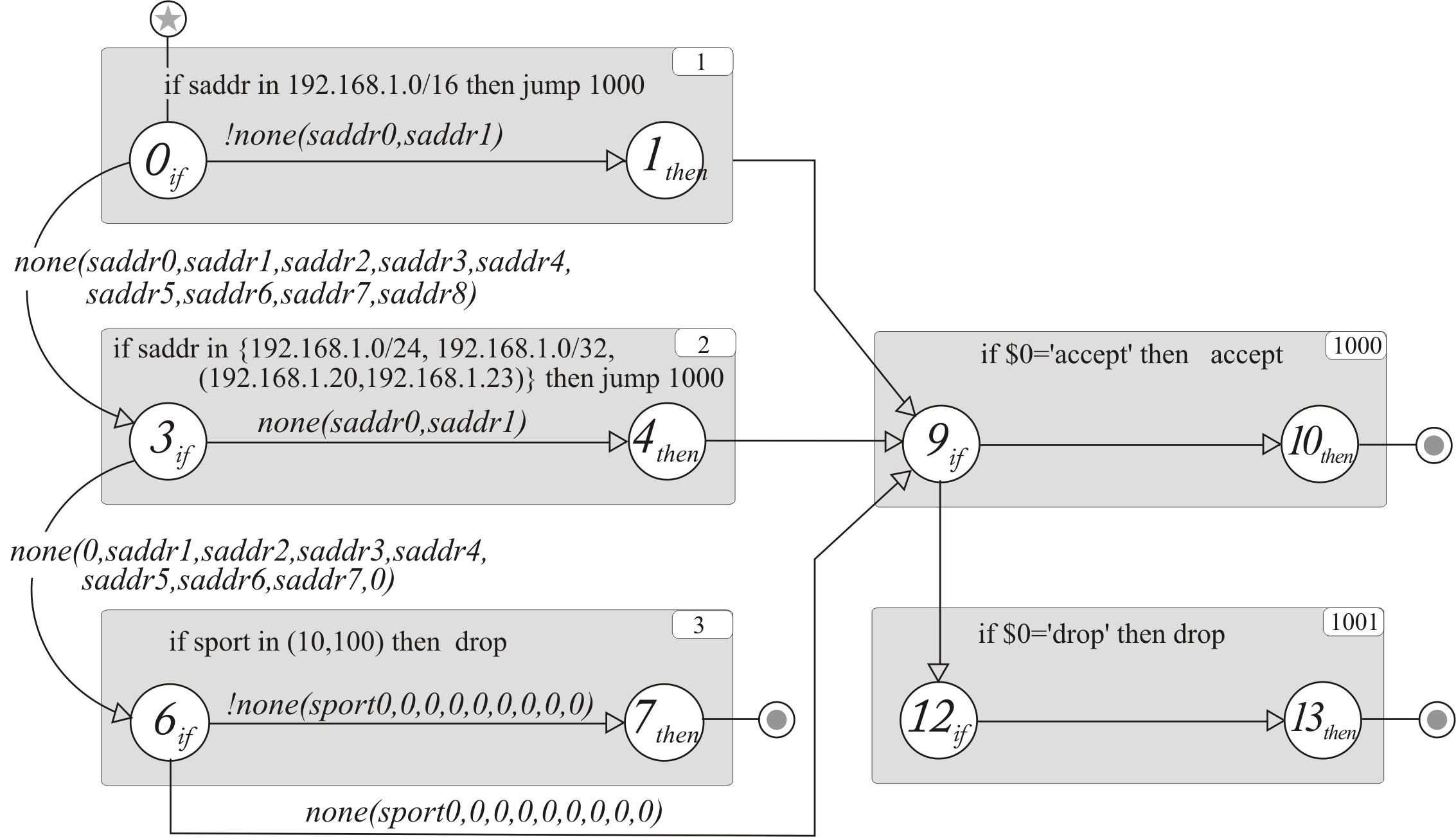}
\caption{Control flow graph for Example 5}
\label{fig:prog_8}
\end{figure}

The control flow graph for this program is shown in
Figure~\ref{fig:prog_8}.

Analysis detects that label 2 has no effect. This is because $saddr$
range check at label 1 is more general than the check at label 2. Any
packet satisfying the check at label 2 will also satisfy the check at label 1 and
thus control flow for such packets will proceed to label 1000. The rule
with label 2 could be safely removed from the policy.

\clearpage\section{Conclusions and Future Work}
In this article we have demonstrated how static code analysis could be
applied to the problem of firewall policy anomaly detection. The
overall framework which we have developed could now be used to apply
more analyses to policies in our intermediate language.

Our implementation of unreachable code detection and live variable
analysis is ``conservative'': it is not guaranteed to detect all
unreachable codes and all variables which are ``dead'' right after
being assigned. While our analysis is conservative, and might not be
able to detect some anomalies, it is ``safe'' and should not produce
false-positives.

We do not detect some anomalies because we do not try to reason about
values of dynamic variables, and whenever dynamic check is present as
part of the rule, we are assuming that both outcomes are possible. In
fact, some of these variables might have values which will make some
branches unreachable which would not be detected by our code. Better
reasoning about dynamic variable values could be a direction of future
research.

In this work we decided not to deal yet with \emph{CALL/RETURN}
statements. This could be done using inter-procedural analysis
techniques and could be an another direction for future work.

We also plan to develop compilers from actual firewall policy
languages to our intermediate language. This would allow us to apply
this work to real-life policies and to make quantitive measurements of
the number of anomalies detected using this approach.

Another very promising direction of future research is application of
these techniques in a distributed firewall environment where we need
to analyze a group of firewall policies interacting together. In this
case, our approach with intermediate policy language will be
especially useful, since polices could be in different languages.

\newpage

\bibliographystyle{acm}
\bibliography{nonlinpolicy}

\newpage\appendix
\appendixpage
\addappheadtotoc

\section{Source Code}
\label{src}

\subsection{Implementation of MCSI calculation algorithm}
\label{mcsisrc}
\begin{tiny}
\begin{verbatim}
-- |  This module implements calculation of Minimal Combining Sets
-- | @author Vadim Zaliva <lord@crocodile.org>
module MCSI where

import Data.List

-- | Interval constructor
data (Ord a, Show a) => Interval a = Interval 
                                     a  -- ^ lower bound
                                     Bool -- ^ left-closed?
                                     a  -- ^ upper bound
                                     Bool -- ^ right-closed?
                                     deriving (Eq)

-- | Defines sort order for intervals, so they are sorted
-- | by their endpongs first. For convenience only.
instance (Ord a, Show a) => Ord (Interval a) where
    compare (Interval a0 ai0 b0 bi0) (Interval a1 ai1 b1 bi1) = 
        compare (a0, b0, ai0, bi0) (a1, b1, ai1, bi1)

-- | Convenience instance for interval pretty-printing
instance (Ord a, Show a) => Show (Interval a) where
    show (Interval a ai b bi) = (if ai then "[" else "(") ++
                                (show a) ++ "-" 
                                ++ (show b) ++
                                (if bi then "]" else ")")


-- | Check whenever point belongs to given interval
inInterval :: (Ord a, Show a) => Interval a -> a -> Bool
inInterval (Interval a ai b bi) x = ((ai && (x>=a)) || (not ai && (x>a))) &&
                                    ((bi && (x<=b)) || (not bi && (x<b)))

splitBy :: (Ord a, Show a) => Interval a -> a -> Bool -> Bool -> [(Interval a)]
splitBy (Interval a ai b bi) x xi xo = 
    if not (inInterval (Interval a ai b bi) x) ||
           (xi && ai && x==a && xo) || 
           (xi && bi && x==b && (not xo)) ||
           a == b
    then
           []
    else if xi then
         nub [norm (Interval a ai x (not xo)), 
              norm (Interval x xo b bi)]
         else
         nub [norm (Interval a ai x False), 
              norm (Interval x True x True), 
              norm (Interval x False b bi)]
    where
    norm (Interval a ai b bi) = if a==b && ai/=bi then (Interval a True b True)
                                else (Interval a ai b bi)

-- | Given set of Intervals calculates MCSI
mcsi :: (Ord a, Show a) => [(Interval a)] -> [(Interval a)]
mcsi il =  let splits = map (splitByList (intervalBounds il)) il
               in if all ((==) []) splits then il
                  else mcsi (nub (splitOrOrig splits il))
    where    
    intervalBounds intlist = nub (foldr1 (++) [ [(a,ai,True),(b,bi,False)] | (Interval a ai b bi) <- intlist])
    splitByList lp i = nub (foldr1 (++) [(splitBy i p pi po) | (p,pi,po)<-lp])
    splitOrOrig (s:xs) (o:os) = if s == [] then 
                                (o : (splitOrOrig xs os))
                                else 
                                s ++ (splitOrOrig xs os)
    splitOrOrig [] [] = []

\end{verbatim}
\end{tiny}

\subsection{Parser Module Sources}
\label{parsersrc}
\begin{tiny}
\begin{verbatim}
{
module PolicyLang where

import Char
import Maybe
import Text.Regex.Posix.String
import Data.Bits
import Data.Word
import Data.List
import System.IO.Unsafe(unsafePerformIO)

import MCSI
import NetworkData
import Policy
}

%name parse
%tokentype { Token }
%monad { P } { thenP } { returnP }
%lexer { lexer } { TokenEOF }

%token 
  if	    { TokenIf }
  then	    { TokenThen }
  nil		{ TokenNil }
  accept	{ TokenAccept }
  drop	    { TokenDrop }
  call	    { TokenCall }
  jump	    { TokenJump }
  return	{ TokenReturn }
  and       { TokenAnd }
  true      { TokenTrue }
  saddr       { TokenSrcAddr }
  sport       { TokenSrcPort }
  daddr       { TokenDstAddr }
  dport       { TokenDstPort }
  in        { TokenIn }
  proto     { TokenProto }
  '!'	    { TokenNeg }
  '='	    { TokenEq }
  '&'       { TokenMask }
  '{'		{ TokenOCB }
  '}'		{ TokenCCB }
  '('		{ TokenORB }
  ')'		{ TokenCRB }
  '['		{ TokenOSB }
  ']'		{ TokenCSB }
  ','		{ TokenComma }
  ';'		{ TokenSemi }
  '/'		{ TokenSlash }
  ':'		{ TokenColumn }
  int		{ TokenInt $$ }
  varname	{ TokenVarName $$ }
  ipv4	    { TokenIPv4 $$ }
  opaque_value	{ TokenOpaqueValue $$ }

%%

ruleset : {- empty -} { [] }
        | ruleset rule  {$2:$1}

rule : int if filteringspec then targetspec ';' { PolicyRule $1 $3 $5} 

targetspec 
    : action_target {$1}
    | call_target  {$1}
    | jump_target  {$1}
    | return_target  {$1}
    | set_target {$1}

action_target: 
      accept { Accept}
      | drop { Drop }

call_target : call int { Call $2 }

jump_target : jump int { Jump $2 }

return_target : return { Return }

set_target : varname '=' opaque_value {Set $1 $3}
     
filteringspec 
   : staticchecks and dynamicchecks {FilteringSpec $1 $3}
   | staticchecks {FilteringSpec $1 DynamicAny}
   | dynamicchecks {FilteringSpec staticAny $1}
   | true {FilteringSpec staticAny DynamicAny}

dynamicchecks : maybeneg varname '=' valuecheck {VarEQ $1 $2 $4}

valuecheck 
    : opaque_value { OpaqueValue $1}
    | nil {IsNil}
    | int { IntValue $1}
    | int '&' int { IntAndMask $1 $3}

maybeneg : '!' { True }
    | { False}

staticchecks : maybeneg saddr_check
               maybeneg sport_check
               maybeneg daddr_check
               maybeneg dport_check
               maybeneg proto_check { StaticCheck $1 $2 $3 $4 $5 $6 $7 $8 $9 $10}

saddr_check: {- empty -} { [] }
           | saddr in addr_interval_or_set {$3}

sport_check: {- empty -} { [] }
           | sport in port_interval_or_set {$3}

daddr_check: {- empty -} { [] }
           | daddr in addr_interval_or_set {$3}

dport_check: {- empty -} { [] }
           | dport in port_interval_or_set {$3}

proto_check: {- empty -} { [] }
           | proto in proto_interval_or_set {$3}

addr_interval_or_set
  : addr_interval { [$1] }
  | '{' addr_interval_set '}' { $2 }

port_interval_or_set
  : port_interval { [$1] }
  | '{' port_interval_set '}' { $2 }

proto_interval_or_set
  : proto_interval { [$1] }
  | '{' proto_interval_set '}' { $2 }

addr_interval:
intrv_open ipv4 ',' ipv4 intrv_close { Interval (readIPv4 $2) $1 (readIPv4 $4) $5 }
| ipv4 '/' int { networkToInterval (CIDR (readIPv4 $1) (read (show $3)))  }
| ipv4 ':' ipv4 { networkToInterval (IPMask (readIPv4 $1) (readIPv4 $3))  }

port_interval: intrv_open int ',' int intrv_close { Interval $2 $1 $4 $5 }

proto_interval: intrv_open int ',' int intrv_close { Interval $2 $1 $4 $5 }

intrv_open
    : '[' {True }
    | '(' { False}

intrv_close
    : ']' {True }
    | ')' { False}

addr_interval_set
    : addr_interval   { [$1] }
    | addr_interval_set ',' addr_interval  { $3 : $1 }

port_interval_set
    : port_interval   { [$1] }
    | port_interval_set ',' port_interval  { $3 : $1 }

proto_interval_set
    : proto_interval   { [$1] }
    | proto_interval_set ',' proto_interval  { $3 : $1 }


{

parse :: P [PolicyRule]

happyError :: P a
happyError = \s i -> error (
  "Parse error in line " ++ show (i::Int) ++ "\n")

data Token
  = TokenIf
  | TokenThen
  | TokenNil
  | TokenAccept
  | TokenDrop
  | TokenCall
  | TokenJump
  | TokenReturn
  | TokenAnd
  | TokenTrue
  | TokenIn
  | TokenSrcAddr
  | TokenDstAddr
  | TokenSrcPort
  | TokenDstPort
  | TokenProto
  | TokenNeg
  | TokenEq
  | TokenMask
  | TokenORB
  | TokenCRB
  | TokenOSB
  | TokenCSB
  | TokenOCB
  | TokenCCB
  | TokenComma
  | TokenSemi
  | TokenSlash
  | TokenColumn
  | TokenInt Int
  | TokenVarName Int
  | TokenIPv4 String
  | TokenOpaqueValue String
  | TokenEOF
  deriving (Show,Eq)
data ParseResult a
  = ParseOk a
  | ParseFail String

type P a = String -> Int -> ParseResult a

thenP :: P a -> (a -> P b) -> P b
m `thenP` k = \s l -> 
  case m s l of
  	ParseFail s -> ParseFail s
  	ParseOk a -> k a s l

returnP :: a -> P a
returnP a = \s l -> ParseOk a

failP :: String -> P a
failP err = \s l -> ParseFail err




lexer :: (Token -> P a) -> P a
lexer cont s = if (length s) == 0 then cont TokenEOF []
               else case skipTokens [skipWS, skipComment] s of
                        Just rest -> lexer cont rest
                        Nothing -> case applyTokenParsers 
                                   [parseFixedToken, 
                                    parseVarName, 
                                    parseIPv4, 
                                    parseOpaque, 
                                    parseInt] s of
                                   Just (token, rest) -> cont token rest
                                   Nothing -> failP "Unknown token" s

applyTokenParsers :: [String -> Maybe (Token, String)] -> String -> Maybe (Token, String)
applyTokenParsers [] _ = Nothing
applyTokenParsers (p:ps) s = case p s of
                             Just (token, rest) -> Just (token, rest)
                             Nothing -> applyTokenParsers ps s


skipTokens :: [(String -> Maybe String)] -> String -> Maybe String
skipTokens [] _ = Nothing
skipTokens (p:ps) s = case p s of 
                      Just rest -> Just rest
                      Nothing -> skipTokens ps s


skipComment :: String -> Maybe String
skipComment "" = Nothing
skipComment (p:ps) = if p == '#' then
                     case findIndex (\c -> c=='\r' || c=='\n') ps of
                                       Nothing -> Just ""
                                       Just n -> Just (drop n ps)
                     else Nothing

                

skipWS :: String -> Maybe String
skipWS "" = Nothing
skipWS s = case findIndex notWS s of
            Nothing -> Just ""
            Just n -> if n==0 then Nothing else Just (drop n s)
    where
    notWS :: Char -> Bool
    notWS c = c/=' ' && c/='\t' && c/='\r' && c/='\n'


parseVarName :: String -> Maybe (Token, String)
parseVarName s = case checkPattern s "\\$[0-9]+" of
                 ("",_) -> Nothing
                 (match,after) -> 
                     Just (TokenVarName (read (tail match)), after)

parseInt :: String -> Maybe (Token, String)
parseInt s = case checkPattern s "[0-9]+" of
             ("",_) -> Nothing
             (match,after) -> 
                 Just (TokenInt (read match), after)

parseOpaque :: String -> Maybe (Token, String)
parseOpaque s = case checkPattern s "'[^']*'" of
                 ("",_) -> Nothing
                 (match,after) -> 
                     Just (TokenOpaqueValue (take ((length match)-2)
                                             (tail match)), after)

parseIPv4 :: String -> Maybe (Token, String)
parseIPv4 s = case checkPattern s "([0-9]){1,3}(\\.[0-9]{1,3}){3}" of
                 ("",_) -> Nothing
                 (match,after) -> 
                     Just (TokenIPv4 match, after)


checkPattern :: String -> String -> (String, String)
checkPattern s p = unsafePerformIO $ do 
                   r <- compile coptions eoptions ("^"++p)
                   case r of
                          Left _ -> return ("", s)
                          Right rc -> 
                              do m <- regexec rc s
                                 case m of
                                        Left _ -> return ("",s)
                                        Right (Just (_, match, after, _)) ->
                                            return (match, after)
                                        Right Nothing -> return ("", s)
    where 
    coptions = compExtended -- IMPORTANT: no compNewline!
    eoptions = execBlank
               
parseFixedToken :: String -> Maybe (Token, String)
parseFixedToken s = parseFixedToken_ tokenDefs s
    where
    tokenDefs :: [(String, Token)]
    tokenDefs = [("if", TokenIf), 
                 ("then", TokenThen), 
                 ("nil", TokenNil),
                 ("accept", TokenAccept),
                 ("drop", TokenDrop),
                 ("call", TokenCall),
                 ("jump", TokenJump),
                 ("return", TokenReturn),
                 ("and", TokenAnd),
                 ("true", TokenTrue),
                 ("in", TokenIn),
                 ("saddr", TokenSrcAddr),
                 ("sport", TokenSrcPort),
                 ("daddr", TokenDstAddr),
                 ("dport", TokenDstPort),
                 ("proto", TokenProto),
                 ("!", TokenNeg),
                 ("=", TokenEq),
                 ("&", TokenMask),
                 ("{", TokenOCB),
                 ("}", TokenCCB),
                 ("(", TokenORB),
                 (")", TokenCRB),
                 ("[", TokenOSB),
                 ("]", TokenCSB),
                 (",", TokenComma),
                 (";", TokenSemi),
                 ("/", TokenSlash),
                 (":", TokenColumn)
                ]
    parseFixedToken_ :: [(String, Token)] -> String -> Maybe (Token, String)
    parseFixedToken_ [] _ = Nothing                          
    parseFixedToken_ _ "" = Nothing                          
    parseFixedToken_ ((p,t):ds) s = if isPrefixOf p s then Just (t, drop (length p) s)
                                    else parseFixedToken_ ds s

------- debugging tools ------

runParser :: String -> [PolicyRule]
runParser s = case parse s 1 of
  	ParseOk e -> reverse e
  	ParseFail s -> error s


testLexer :: String -> [Token]
testLexer s = if (length s) == 0 then []
              else 
              case skipWS s of
                            Just rest -> testLexer rest
                            Nothing -> case applyTokenParsers 
                                       [parseFixedToken, 
                                        parseVarName, 
                                        parseIPv4, 
                                        parseOpaque, 
                                        parseInt] s of
                                       Just (token, rest) -> token:(testLexer rest)
                                       Nothing -> [TokenEOF]

}
\end{verbatim}
\begin{verbatim}
{--
  Definition of basic networking data types, IP addresses, networks
  ranges.

  Currently working with IPv4 addresses only.
--}

module NetworkData where

import Data.Bits
import Data.Word
import Data.List
import MCSI

type Octet = Word
data IP = IP Octet Octet Octet Octet
          deriving (Eq)
type IPMask = IP

data Network = IPMask IP IPMask | CIDR IP Word8

instance Bounded IP where
    minBound = IP 0 0 0 0
    maxBound = IP 255 255 255 255

instance Ord IP where
    compare a b =
        compare (ip2word a) (ip2word b)

instance Show IP where
    show (IP a b c d) = (show a) ++ "." 
                         ++ (show b) ++ "." 
                         ++ (show c) ++ "."
                         ++ (show d)

ip2word :: IP -> Word32
ip2word (IP a0 b0 c0 d0) = fromInteger (
    (toInteger d0)+(toInteger c0)*256+(toInteger b0)*(256^2)+(toInteger a0)*(256^3))

word2ip :: Word32 -> IP
word2ip x = IP w1 w2 w3 w4
            where
            (o1,o1m) = divMod x (256^3)
            (o2,o2m) = divMod o1m (256^2)
            (o3,o3m) = divMod o2m 256
            o4 = mod x 256
            w1 = fromInteger (toInteger o1)
            w2 = fromInteger (toInteger o2)
            w3 = fromInteger (toInteger o3)
            w4 = fromInteger (toInteger o4)

instance Show Network where
    show (IPMask ip mask) = (show ip) ++ ":" ++ (show mask)
    show (CIDR ip prefix) = (show ip) ++ "/" ++ (show prefix)

type IPInterval = Interval IP

split delim s
    | [] <- rest = [token]
    | otherwise = token : split delim (tail rest)
    where (token,rest) = span (/=delim) s

readIPv4 :: String -> IP
readIPv4 s = let l = split '.' s  in
                     IP (read (l!!0)) (read (l!!1)) (read (l!!2)) (read (l!!3))

prefix2mask :: Word8 -> Word32
prefix2mask prefix = prefix2mask_ prefix 0
    where
    prefix2mask_:: Word8 -> Word32 -> Word32
    prefix2mask_ 0 x = x
    prefix2mask_ p x = prefix2mask_ (p-1) (setBit (shiftR x 1) 31)

networkToInterval:: Network -> IPInterval
networkToInterval (IPMask ip mask) = 
    Interval
    (word2ip ((ip2word ip) .&. (ip2word mask)))
    True
    (word2ip ((ip2word ip) .|. (complement (ip2word mask))))
    True
networkToInterval (CIDR ip prefix) = 
    networkToInterval (IPMask ip (word2ip (prefix2mask prefix)))

universalIPInterval:: IPInterval
universalIPInterval = Interval minBound True maxBound True

universalPortInterval:: Interval Int
universalPortInterval = Interval 0 True 65535 True

universalProtoInterval:: Interval Int
universalProtoInterval = Interval 0 True 255 True

\end{verbatim}
\begin{verbatim}
{--
  Definition firewall policy related data types
--}


module Policy where
import Data.Word
import MCSI
import NetworkData

type RuleLabel = Int -- Rule #
type VarName = Int -- Variable name/#
type VarValue = String -- Opaque variable value
type Neg = Bool -- Negation flag
type Port = Int -- TCP Port #
type Protocol = Int -- IP Protocol #

data PolicyRule = PolicyRule RuleLabel FilteringSpec Target
                deriving(Show, Eq)

data FilteringSpec = FilteringSpec StaticCheck DynamicCheck
                deriving(Show, Eq)

data DynamicCheck =  VarEQ Neg VarName DynamicCheckValue
                  | DynamicAny
                  deriving(Show, Eq)

data DynamicCheckValue = 
    OpaqueValue VarValue 
    | IntValue Int 
    | IntAndMask Int Int
    | IsNil
      deriving(Show, Eq)

data Target = Accept |
              Drop | 
              Call RuleLabel |
              Jump RuleLabel |
              Return |
              Set VarName VarValue
                  deriving(Show, Eq)


data StaticCheck = StaticCheck Neg [(Interval IP)] Neg [(Interval Port)] Neg [(Interval IP)] Neg [(Interval Port)] Neg [(Interval Protocol)] 
                   deriving(Show, Eq)


staticAny :: StaticCheck 
staticAny = StaticCheck False [] False [] False [] False [] False []

\end{verbatim}
\end{tiny}

\subsection{Data Flow and CFG Extractor Module Sources}
\label{transformsrc}
\begin{tiny}
\begin{verbatim}
{--
  Policy transformations
-}

module PolicyTransform(xmain) where 

import System( getArgs )
import IO
import Data.List(nub,sort,zip5,elemIndex)
import Ix(range)

import MCSI
import Policy
import PolicyLang
import NetworkData

type InternalLabel = Int -- Internal Rule label

-- TODO add unit tests.
combineFromMCSI :: (Ord a, Show a) => [(Interval a)] -> [(Interval a)] -> [(Interval a)]
combineFromMCSI domain old = filter (\m -> any (inside m) old) domain
    where
    inside (Interval x xi y yi) h = inside1 h x xi True && inside1 h y yi False
    inside1 (Interval a ai b bi) x xi xo = (inInterval (Interval a ai b bi) x )
                                           || ((not xi) && (not ai) && x==a && xo) 
                                           || ((not xi) && (not bi) && x==b && (not xo)) 


getSaddr (PolicyRule _ (FilteringSpec (StaticCheck n x _ _ _ _ _ _ _ _) _) _) = (n,x)
getSport (PolicyRule _ (FilteringSpec (StaticCheck _ _ n x _ _ _ _ _ _) _) _) = (n,x)
getDaddr (PolicyRule _ (FilteringSpec (StaticCheck _ _ _ _ n x _ _ _ _) _) _) = (n,x)
getDport (PolicyRule _ (FilteringSpec (StaticCheck _ _ _ _ _ _ n x _ _) _) _) = (n,x)
getProto (PolicyRule _ (FilteringSpec (StaticCheck _ _ _ _ _ _ _ _ n x) _) _) = (n,x)

getLabel (PolicyRule x _ _) = x


policyLabels:: [PolicyRule] -> [RuleLabel]
policyLabels [] = []
policyLabels ((PolicyRule l _ a):xs) = case a of 
                                              (Call m) -> [l,m] ++ (policyLabels xs)
                                              (Jump m) -> [l,m] ++ (policyLabels xs)
                                              _ -> l:(policyLabels xs)


-- Maps internal to original label
internalToOrig :: [PolicyRule] -> InternalLabel -> RuleLabel
internalToOrig rules l = (nub (sort (policyLabels rules))) !! (floor ((fromIntegral l)/3))

-- Maps policy label from original policy to pair of internal sublabels
-- First one is for condition and the second one is for 'then' branch,
-- and the third one for return from 'call'.
-- The mapping preserves the order.
label2internal :: [PolicyRule] -> RuleLabel -> (InternalLabel,InternalLabel,InternalLabel)
label2internal policy l = case elemIndex l (nub (sort (policyLabels policy))) of
                    Nothing -> (-1,-1,-1)
                    Just pos -> (pos*3, pos*3+1, pos*3+2)

-- Returns internal label for 'if' part of original label
label2internalIf :: [PolicyRule] -> RuleLabel -> InternalLabel
label2internalIf policy l = let (a,_,_) = label2internal policy l
                                in a

-- Returns internal label for 'then' part of original label
label2internalThen :: [PolicyRule] -> RuleLabel -> InternalLabel
label2internalThen policy l = let (_,a,_) =  label2internal policy l
                                  in a

-- Returns internal label for 'return' part of original label
label2internalRet :: [PolicyRule] -> RuleLabel -> InternalLabel
label2internalRet policy l = let (_,_,a) =  label2internal policy l
                                  in a

internalLabels:: [PolicyRule] -> [InternalLabel]
internalLabels rules =  
    let all_succ = rules2succ rules 
        in
        sort $ nub (
        foldl1 (++)  [ [i0,i1] | (i0,i1) <- (foldl1 (++) [ u++t++f | (_,u,t,f) <- all_succ])]
                  ++
        foldl1 (++) [ [label2internalIf rules l] | (l,_,_,_) <- all_succ])
        

labelsPedicate :: [PolicyRule] -> String
labelsPedicate rules =  foldl (++) "" [ "label("++(show r)++").\n" | r<-internalLabels rules ]

olabelsPedicate :: [PolicyRule] -> String
olabelsPedicate rules =  foldl (++) "" [ "olabel("++(show (internalToOrig rules l))++","++(show l)++").\n" | l<-internalLabels rules ]
    


initPredicate :: [PolicyRule] -> String
initPredicate rules = "init(" ++ show (minimum (internalLabels rules)) ++ ").\n"

policyVariables:: [PolicyRule] -> [VarName]
policyVariables [] = []
policyVariables ((PolicyRule _ (FilteringSpec _ dcheck) tspec):xs) = 
    tl tspec ++ fl dcheck ++ policyVariables xs
    where
    tl (Set v _) = [v]
    tl _ = []
    fl (VarEQ _ v _) = [v]
    fl _ = []

renumberVariables :: [PolicyRule] -> [PolicyRule]
renumberVariables rules = map (renumber (sort $ nub $ policyVariables rules)) rules
    where
    renumber vars (PolicyRule l (FilteringSpec scheck dcheck) tspec) =
        (PolicyRule l (FilteringSpec scheck (renD vars dcheck)) (renT vars tspec))
    renT vars (Set v x) = (Set (mapvars vars v) x)
    renT _ x = x
    renD vars (VarEQ x v y) = (VarEQ x (mapvars vars v) y)
    renD _ x = x
    mapvars vars v =
        case elemIndex v vars of
                              Nothing -> -1
                              Just pos -> pos

varPedicate :: [PolicyRule] -> String
varPedicate rules =  foldl (++) "" [ "var("++(show r)++").\n" | r<-(sort $ nub $ policyVariables rules) ]

varReads:: [PolicyRule] -> [(RuleLabel,VarName)]
varReads [] = []
varReads (x:xs) = 
    case x of
           (PolicyRule l (FilteringSpec _ (VarEQ _ v _)) _) -> (l,v):(varReads xs)
           _ -> varReads xs

varReadsInternal:: [PolicyRule] -> [(InternalLabel,VarName)]
varReadsInternal rules = [((label2internalIf rules l), n)  | (l,n)<-(varReads rules)]

readPedicate :: [PolicyRule] -> String
readPedicate rules =  foldl (++) "" [ "read("++(show l)++","++ (show v)++").\n" | (l,v)<-(sort $ nub $ varReadsInternal rules) ]

varWrites:: [PolicyRule] -> [(RuleLabel,VarName)]
varWrites [] = []
varWrites (x:xs) = 
    case x of
           (PolicyRule l (FilteringSpec _ _) (Set v _)) -> (l,v):(varWrites xs)
           _ -> varWrites xs

varWritesInternal:: [PolicyRule] -> [(InternalLabel,VarName)]
varWritesInternal rules = [((label2internalThen rules l), n)  | (l,n)<-(varWrites rules)]

writePedicate :: [PolicyRule] -> String
writePedicate rules =  foldl (++) "" [ "write("++(show l)++","++ (show v)++").\n" | (l,v)<-(sort $ nub $ varWritesInternal rules) ]

finals:: [PolicyRule] -> [PolicyRule] -> [RuleLabel]
finals rules [] = []
finals rules ((PolicyRule l _ Accept):xs) = (label2internalThen rules l):(finals rules xs)
finals rules ((PolicyRule l _ Drop):xs) = (label2internalThen rules l):(finals rules xs)
finals rules ((PolicyRule l _ (Set _ _)):[]) = [(label2internalThen rules l), (label2internalIf rules l)]
finals rules ((PolicyRule l _ _):xs) = finals rules xs

finalsPedicate :: [PolicyRule] -> String
finalsPedicate rules =  foldl (++) "" [ "final("++(show r)++").\n" | r<-(finals rules rules)]

buildChecks sadom spdom dadom dpdom pdom ruleset = 
    [[("saddr", isNeg (getSaddr x), (mapM (getSaddr x) sadom)),
      ("sport", isNeg (getSport x), (mapM (getSport x) spdom)),
      ("daddr", isNeg (getDaddr x), (mapM (getDaddr x) dadom)),
      ("dport", isNeg (getDport x), (mapM (getDport x) dpdom)),
      ("proto", isNeg (getProto x), (mapM (getProto x) pdom ))]
        | x<-ruleset]
    where
    isNeg (n,_) = n
    mapM (_,ls) dom = [ elem d (combineFromMCSI dom ls) | d<-dom]



noneArity sadom spdom dadom dpdom pdom = maximum [length sadom,
                                                  length spdom,
                                                  length dadom,
                                                  length dpdom,
                                                  length pdom]

nonePedicate sadom spdom dadom dpdom pdom = 
    let n = noneArity sadom spdom dadom dpdom pdom 
        in 
        ((noneProto n), (noneDef n))
    where
    noneProto n =  "none("++(commaSeparated [ "b"++(show x)++":B"| x<- take n (iterate (1+) 1)]) ++")\n"
    noneDef n =  "none("++(commaSeparated [ "x"++(show x)| x<- take n (iterate (1+) 1)]) ++") :- " ++
                 (commaSeparated [ "x"++(show x)++"=0"| x<- take n (iterate (1+) 1)]) ++ ".\n"
                                      



rule2succ :: [PolicyRule] -> PolicyRule -> (Maybe PolicyRule) -> 
             (RuleLabel, [(InternalLabel,InternalLabel)],[(InternalLabel,InternalLabel)],[(InternalLabel,InternalLabel)])
rule2succ allrules r0 r1 = 
    let (a0,b0,c0) = part1 allrules r0
        (a1,b1,c1) = part2 allrules r0 r1
        l = getLabel r0
        in
        (l, a0++a1,b0++b1,c0++c1)
    where
    part1 allrules (PolicyRule l0 _ t) = 
        let l0if = label2internalIf allrules l0
            l0then = label2internalThen allrules l0
            in
            case t of 
                   (Jump lx) -> ([(l0then, (label2internalIf allrules lx))],
                                 [(l0if, l0then)],
                                 [])
                   (Call lx) -> ([(l0then,(label2internalIf allrules lx))],
                                 [(l0if, l0then)],
                                 [])
                   (Set _ _) -> ([],
                                 [(l0if,l0then)],
                                 [])
                   (Accept) -> ([],
                                [(l0if,l0then)],
                                [])
                   (Drop) -> ([],
                              [(l0if,l0then)],
                              [])
                   _ -> ([],[],[])

    part2 allrules (PolicyRule l0 _ t) r1 = 
        case r1 of
                Nothing -> ([],[],[])
                Just (PolicyRule l1 _ _) ->  
                    let 
                    l0if = label2internalIf allrules l0
                    l0then = label2internalThen allrules l0
                    l0ret = label2internalRet allrules l0
                    l1if = label2internalIf allrules l1
                    l1then = label2internalThen allrules l1
                    in
                    case t of 
                           (Jump lx) -> ([],
                                         [],
                                         [(l0if, l1if)])
                           (Call lx) -> ([(l0ret,l1if)],
                                         [],
                                         [(l0if, l1if)])
                           (Set _ _) -> ([(l0then, l1if)],
                                         [],
                                         [(l0if, l1if)])
                           (Accept) -> ([],
                                        [],
                                        [(l0if, l1if)])
                           (Drop) -> ([],
                                      [],
                                      [(l0if, l1if)])
                           _ -> ([],[],[])

rules2succ :: [PolicyRule] -> [(RuleLabel,[(InternalLabel,InternalLabel)],[(InternalLabel,InternalLabel)],[(InternalLabel,InternalLabel)])]
rules2succ rules = rules2succ_ rules rules
    where
    rules2succ_ rules [] = []
    rules2succ_ rules (x:[]) = [rule2succ rules x (Nothing)]
    rules2succ_ rules (x:(y:xs)) = (rule2succ rules x (Just y)):(rules2succ_ rules (y:xs))

succPedicate sadom spdom dadom dpdom pdom rules = 
    let all_c = buildChecks sadom spdom dadom dpdom pdom rules
        all_succ = rules2succ rules 
        narity = noneArity sadom spdom dadom dpdom pdom
        in
                foldl (++) "" [ "# Label " ++ (show l) ++
                                foldl (++) "\n#  unconditional:\n" (map (bsu c) u) ++
                                foldl (++) "#  if condidion satified\n" (map (bs c narity) t) ++ 
                                foldl (++) "#  if condidion not satified\n" (map (bsn c narity) f)
                                | ((l,u,t,f),c) <- zip all_succ all_c]
    where
    -- unconditional clause
    bsu c (l, x) = (pproto l x c) ++ ".\n"

    -- positive clause
    {-- positive clause. if no static variables involved, generates unconditional
      succ() preducate in from succ(a,b,_,_...). -}
    bs c narity (l, x) = if or [or vflags | (_, _, vflags)<-c] then
                         (pproto l x c) ++ " :- " ++ (commaSeparated (pslist c narity)) ++ ".\n"
                         else (pproto l x c) ++ ".\n"



    -- negative clause
    bsn c narity (l, x) = if or [or vflags | (_, _, vflags)<-c] then
                          foldl (++) "" [ (bs_ (ignoreothers n c) (l, x)) | n <- range (0 ,(length c)) ]
                          else (pproto l x c) ++ ".\n"
                              where
                              bs_ c (l, x) = if or [or vflags | (_, _, vflags)<-c] then
                                             (pproto l x c) ++ " :- " ++ (commaSeparated (pslist c narity)) ++ ".\n"
                                             else ""


    -- negates one constraint clause and ignores others
    ignoreothers :: Int -> [(String,Bool,[Bool])] -> [(String,Bool,[Bool])]
    ignoreothers _ [] = []
    ignoreothers n ((vname, vneg, vflags):cs) = 
        if n == 0 then (vname, (not vneg), vflags):(ignoreothers (n-1) cs)
        else (vname, vneg, (take (length vflags) (repeat False))):(ignoreothers (n-1) cs)

    pslist :: [(String,Bool,[Bool])] -> Int -> [String]
    pslist [] _ = []
    pslist ((vname, vneg, vflags):xs) narity = 
        if or vflags then (callNone vneg (varslist 0 vname vflags) narity):(pslist xs narity)
        else pslist xs narity
        
    callNone neg ls narity = (if neg then "" else "!") ++
                      "none(" ++ (commaSeparated (ls ++ (replicate (narity - (length ls)) "0"))
                                 ) ++ ")" 

    varslist :: Int -> String -> [Bool] -> [String]
    varslist _ _ [] = []
    varslist n prefix (x:xs) = if x then  (p prefix n x):(varslist (n+1) prefix xs)
                               else "0":(varslist (n+1) prefix xs)

    pproto :: InternalLabel -> InternalLabel -> [(String, Bool, [Bool])] -> String
    pproto l0 l1 c = "succ("++(show l0)++","++ (show l1)++","++(plist c)++")"

    plist :: [(String, Bool, [Bool])] -> String
    plist c = commaSeparated (foldl (++) [] [varlist 0 vname vflags | (vname, _, vflags)<-c])

    varlist :: Int -> String -> [Bool] -> [String]
    varlist _ _ [] = []
    varlist n prefix (x:xs) = (p prefix n x):(varlist (n+1) prefix xs)

    p :: String -> Int -> Bool -> String
    p name n True = name ++ (show n)
    p _ _ False = "_"


commaSeparated:: [String] -> String
commaSeparated [] = ""
commaSeparated l = foldl1 (\a b -> a ++ "," ++ b) l

stdProto :: String
stdProto = 
    "olabel(o0:O,l0:L) input\n" ++
    "label(l0:L) input\n" ++
    "init(l:L) input\n" ++
    "final(l:L) input\n" ++
    "var(v:V) input\n" ++
    "write(l:L,v:V) input\n" ++
    "read(l:L,v:V) input\n" ++
    "rechable(l:L) outputtuples\n" ++
    "unrechable(l:L) outputtuples\n" ++
    "suspect(l:L) outputtuples\n" ++
    "osuspect(o:O) outputtuples\n" ++
    "live(l:L,v:V) outputtuples\n" ++
    "dead(l:L) outputtuples\n"

dynamic sadom spdom dadom dpdom pdom =
    let n = sum [length sadom,
                 length spdom,
                 length dadom,
                 length dpdom,
                 length pdom]
        proto = commaSeparated [ "b"++(show x)++":B"| x<- take n (iterate (1+) 1)]
        args = commaSeparated [ "x"++(show x)| x<- take n (iterate (1+) 1)]
        in 
        (genproto proto, gendef args)
    where
    genproto p =
        "path(l0:L,l1:L, " ++ p ++ ")\n" ++
         "readonlyPath(l0:L,l1:L,v:V, " ++ p ++ ")\n" ++
        "succ(l0:L,l1:L, " ++ p ++ ") input\n"
    gendef a =
        "rechable(L) :- init(Li), path(Li,L,"++a++").\n"++
        "unrechable(L) :- label(L), !rechable(L).\n" ++
        "live(L,V) :- init(Li), path(Li,L,"++a++"), read(Lr,V), readonlyPath(L,Lr,V,"++a++").\n"++
        "dead(L) :- write(L,V), !live(L,V).\n"++
        "suspect(L) :- dead(L).\n"++
        "osuspect(O) :- olabel(O,X), suspect(X).\n" ++
        "suspect(L) :- unrechable(L).\n"++
        "path(A,B,"++a++") :- A=B, label(A), label(B).\n"++
        "path(A,C,"++a++") :- succ(A,B,"++a++"), path(B,C,"++a++"), label(A), label(B), label(C).\n"++
        "readonlyPath(A,B,_,"++a++") :- A=B, label(A), label(B).\n"++
        "readonlyPath(A,B,_,"++a++") :- succ(A,B,"++a++"), label(A), label(B).\n"++
        "readonlyPath(A,C,V,"++a++") :- readonlyPath(A,B,V,"++a++"), succ(B,C,"++a++"), var(V), label(A), label(B), label(C), !write(B,V).\n"




xmain = do
       args <- getArgs
       file <- readFile (head args)
       let policy = runParser file
           sadom = mcsi (foldl (++) [] [ x | (_, x) <- (map getSaddr policy)])
           spdom = mcsi (foldl (++) [] [ x | (_, x) <- (map getSport policy)])
           dadom = mcsi (foldl (++) [] [ x | (_, x) <- (map getDaddr policy)])
           dpdom = mcsi (foldl (++) [] [ x | (_, x) <- (map getDport policy)])
           pdom  = mcsi (foldl (++) [] [ x | (_, x) <- (map getProto policy)])
           predicates = generatePredicates sadom spdom dadom dpdom pdom policy
           (noneproto, nonedef) = nonePedicate sadom spdom dadom dpdom pdom
           (dproto, ddef) = dynamic sadom spdom dadom dpdom pdom
           domains = generateDomains policy
       hPutStr stdout "\n\n### Domains\n"
       hPutStr stdout domains
       hPutStr stdout "\n\n### Relations\n"
       hPutStr stdout noneproto
       hPutStr stdout stdProto
       hPutStr stdout dproto
       hPutStr stdout "\n\n### Data\n"
       hPutStr stdout predicates
       hPutStr stdout "\n\n### Rules\n"
       hPutStr stdout nonedef
       hPutStr stdout ddef
       hPutStr stdout "\n\n### Goals\n"
       hPutStr stdout "osuspect(X)?\n"
       hPutStr stdout "\n\n### EOF\n"

{-
       hPutStr stdout (foldr (\x y -> (show x) ++ "\n" ++ y) ""  (buildChecks policy))
       hPutStr stdout (foldr (\x y -> (show x) ++ "\n" ++ y) ""  (normalizeIntervals policy))
       hPutStr stdout (foldr (\x y -> (show x) ++ "\n" ++ y) ""  (normalizeIntervals policy))
       hPutStr stdout "\n\n"
       hPutStr stdout (foldr (\x y -> (show x) ++ "\n" ++ y) ""  (buildAssumptions policy))
       hPutStr stdout "\n\n"
       hPutStr stdout (foldr (\x y -> (show x) ++ "\n" ++ y) ""  (buildUnCond policy))
-}


maximumOr0 [] = 0
maximumOr0 x = maximum x

generateDomains :: [PolicyRule] -> String
generateDomains policy = let
                         nlabels = (maximumOr0 (internalLabels policy)) + 1
                         nvars = (maximumOr0 (sort $ nub $ policyVariables policy)) + 1 
                         in
                         "O " ++ (show (1 + (maximumOr0 (policyLabels policy)))) ++ " olabels.map\n" ++
                         "L " ++ (show nlabels) ++ " labels.map\n" ++
                         "V " ++ (show nvars) ++ " vars.map\n" ++
                         "B 2\n"


generatePredicates sadom spdom dadom dpdom pdom policy =  
    let
    vpolicy = renumberVariables policy
    in
    (labelsPedicate policy) ++
    (olabelsPedicate policy) ++
    (finalsPedicate policy) ++
    (varPedicate vpolicy) ++
    (readPedicate vpolicy) ++
    (writePedicate vpolicy) ++
    (initPredicate policy) ++
    (succPedicate sadom spdom dadom dpdom pdom  policy)




\end{verbatim}
\end{tiny}

\end{document}